\batchmode
\makeatletter
\def\input@path{{"C:/Dropbox/Missing Data/arxiv-sub-2024-01/paper-missing-signals/"}}
\makeatother
\documentclass[12pt,english]{article}
\usepackage[T1]{fontenc}
\usepackage[latin9]{inputenc}
\usepackage{verbatim}
\usepackage{amsmath}
\usepackage{setspace}
\usepackage[authoryear]{natbib}

\makeatletter
\usepackage{chenpaper}
\usepackage{bookmark}
\usepackage{longtable}
\usepackage{setspace}

\makeatother

\usepackage{babel}
\begin{document}
\title{\textbf{\color{Black}Missing Values Handling for Machine Learning Portfolios}}
\author{\large {Andrew Y. Chen}\\{\normalsize Federal Reserve Board} \and \large {Jack McCoy}\\{\normalsize Columbia University}}
\date{January 2024\thanks{This paper was previously circulated as ``Missing Values and the
Dimensionality of Expected Returns.'' First submitted to arXiv.org:
July 20, 2022. Emails: andrew.y.chen@frb.gov (Chen, corresponding
author) and jmccoy26@gsb.columbia.edu (McCoy). This project originated
from many conversations with Fabian Winkler. We thank an anonymous
referee, Heiner Beckmeyer, Charlie Clarke, Harry Mamaysky, Markus
Pelger, and Yinan Su for helpful comments. The views expressed herein
are those of the authors and do not necessarily reflect the position
of the Board of Governors of the Federal Reserve or the Federal Reserve
System.}}
\maketitle
\begin{abstract}
\begin{singlespace}
\noindent We characterize the structure and origins of missingness
for 159 cross-sectional return predictors and study missing value
handling for portfolios constructed using machine learning. Simply
imputing with cross-sectional means performs well compared to rigorous
expectation-maximization methods. This stems from three facts about
predictor data: (1) missingness occurs in large blocks organized by
time, (2) cross-sectional correlations are small, and (3) missingness
tends to occur in blocks organized by the underlying data source.
As a result, observed data provide little information about missing
data. Sophisticated imputations introduce estimation noise that can
lead to underperformance if machine learning is not carefully applied.
\end{singlespace}
\end{abstract}
\vspace{10ex}
\noindent \textbf{\color{Black}JEL Classification}: G0, G1

\noindent \textbf{\color{Black}Keywords}: stock market predictability,
stock market anomalies, missing values, machine learning \thispagestyle{empty}\setcounter{page}{0}

\vspace{10ex}

\pagebreak{}

\section{Introduction}

A growing literature applies methods from machine learning (ML) to
asset pricing. These studies combine the information in dozens, or
even hundreds, of cross-sectional stock return predictors. \citet{freyberger2020dissecting}
combine 62; \citet{kelly2023principal} combine 138; and \citet{han2022expected}
combine up to 193 predictors. Each of these papers finds that there
are economic gains to expanding the set of predictors beyond the five
used in \citet{fama2015five}.\footnote{\citet{lewellen2015cross} finds that regressions on 15 predictors
lead to an annualized Sharpe Ratio of around 0.8 (excluding micro-caps).
\citet{freyberger2020dissecting}; \citet{kelly2023principal}; and
\citet{han2022expected} find Sharpe ratios in excess of 1.8, and
as high as 2.4 (excluding micro-caps or value-weighting). Other papers
that combine many predictors include \citet{haugen1996commonality,green2017characteristics,demiguel2020transaction,gu2020empirical,lopez2020common,kozak2020shrinking,chen2022zeroing,simon2022deep,chen2023deep}.}

Buried in this literature is the problem of missing values. When learning
from many predictors, the standard practice of dropping stocks with
missing values is often untenable. For example, applying the standard
practice to the 125 most-observed predictors in the \citet{ChenZimmermann2022}
dataset drops 99\% of stocks. So even though imputing missing values
may seem dangerous, ML researchers often have no choice but to impute.

We examine missing data handling for 159 predictors using a broad
array of imputation and return forecasting methods. We focus on cross-sectional
expectation-maximization (EM) and cross-sectional mean imputation,
but also examine several other methods, including some that incorporate
time-series information. Our baseline forecasts use principal component
regressions (PCR), but we also examine neural networks and gradient
boosting. These methods are chosen to provide an overview and intuition
behind missing data issues, rather than to advocate for a particular
imputation algorithm.

We find that imputing with cross-sectional means (as in \citet*{kozak2020shrinking,gu2020empirical})
does a surprisingly good job of capturing the potential expected returns.
Imputing with means and then sorting stocks on return forecasts from
a three-layer neural network leads to long-short decile returns of
66\% per year equal-weighted or 37\% per year value-weighted. Imputing
with cross-sectional EM leads to nearly identical returns: 67\% equal-weighted
and 39\% value-weighted.\footnote{The returns from PCR in mean-imputed data are 51\% and 32\% (equal-
and value-weighted, respectively), compared to 54\% and 36\%, in EM-imputed
data. All results use the usual assumption of no transaction costs,
including no shorting fees.} This invariance is seen across most forecasting and imputation methods,
despite the fact that EM has strong theoretical properties and is
recommended in textbooks on missing data handling and machine learning
(\citet{little2019statistical,efron2016computer}).

This invariance is intuitive given three properties of cross-sectional
predictor data. The first is that missingness occurs in large blocks
organized by time. For common stocks on major exchanges that are missing
book-to-market this month, 81\% are also missing book-to-market in
every previous month. As a result, there is little time-series information
about missing values.

The second key property is that predictors are largely uncorrelated
cross-sectionally (\citet*{green2013supraview,ChenZimmermann2022}).
Almost all correlations lie between $-0.25$ and $+0.25$, and the
first 10 principal components span only 40\% of total variance. Thus,
most observed predictors contain little cross-sectional information
about the missing predictors.

The third key property is that missingness tends to occur in blocks
organized by the underlying data. For stocks that are missing book-to-market,
about 90\% are also missing earnings-to-price. So even though book-to-market
and earnings-to-price have a nontrivial cross-sectional correlation
of around 0.3, this correlation can rarely be used for imputing book-to-market.

Overall, we find the observed predictors provide little information
about the missing predictors, so one might as well impute with cross-sectional
means.

In some settings, EM imputation can even lead to \emph{under}performance.
Our baseline results form forecasts separately for micro-cap, small-cap,
and big stocks, motivated by the fact that predictability can differ
by market cap and market liquidity (\citet{fama2008dissecting,chen2022zeroing}).
If we instead form forecasts using all stocks simultaneously, EM often
leads to smaller value-weighted returns compared to simple mean imputation,
though equal-weighted returns are generally similar.

This underperformance is consistent with the idea that sophisticated
imputations introduce estimation noise. This noise can lead to underperformance,
particularly if the forecasts are not carefully structured. Consistent
with this interpretation, we find that EM imputation errors are larger
among small stocks in random masking exercises.

All together, we recommend using simple cross-sectional mean imputation
for handling missing values when applying machine learning to cross-sectional
return predictors. In theory, EM should be less biased, more efficient,
and thus lead to more accurate forecasts. In practice, the missingness
and covariance structures of cross-sectional predictors imply that
the improvements are small, and estimation noise can even lead to
underperformance.

The high dimensionality of cross-sectional predictors naturally raises
the question of how many of these dimensions contribute to expected
returns. Our principal components (PC) regression tests find that
the answer is subtle and depends on multiple measurement choices.
Computing PCs the standard way, we find that about 75 PCs contribute
to the potential equal-weighted returns of 50\% per year, while 40
PCs contribute to the potential value-weighted returns of about 20\%.
However, standard PC analysis is an unsupervised method that ignores
return information. Using \citet{huang2022scaled}'s scaled PCA to
incorporate return information leads to a much smaller dimensionality:
about 30 PCs are required when equal-weighting while 15 are required
when value-weighting.

We also describe the origins of missing values in the \citet{ChenZimmermann2022}
dataset. We find there are three main drivers: (1) missing underlying
data, (2) predictors requiring a long history of underlying data,
and (3) using missingness as a substitute for interaction effects.
Requiring accounting variables leads to an observed share of around
70\% in 1985 and requiring analyst forecasts drops the observed share
to about 50\%. Variables that require a long history of data include
5-year sales growth, which is observed for about 40\% of stocks. Interaction
effects are implicit in predictors like asset tangibility and payout
yield, which are replaced with missing values for non-manufacturing
firms and stocks with non-positive payout, respectively. These drivers
of missingness can themselves interact and lead to extremely low rates
of observability. For example, \citet{asquith2005short}'s institutional
ownership among high short interest stocks predictor requires both
specialized underlying data and uses missingness to represent interaction
effects, leading to a 0.2\% of stocks having observations in 1985.

We post all of our code at https://github.com/jack-mccoy/missing\_data.
The EM imputed data can be found at https://sites.google.com/site/chenandrewy/.

\subsection{Related Literature}

In contemporaneous work, \citet*{freyberger2021missing} (FHNW), \citet*{bryzgalova2022missing}
(BLLP), and \citet{beckmeyer2022recovering} also study missing values
in cross-sectional predictor data. Each paper advocates a different
imputation algorithm, though all avoid modeling the missingness process.
FHNW use moment conditions designed for cross-sectional predictors,
BLLP use \citet{xiong2022large}'s latent factor method, and Beckmeyer
and Wiedemann use a masked language model. Our paper takes a more
neutral approach and compares textbook imputation methods (\citet{little2019statistical})
with the cross-sectional mean imputation used in asset pricing. Our
approach leads us to focus on practical issues, like whether imputation
affects inferences made from standard ML algorithms like neural networks
and gradient boosting (\citet{efron2016computer}). We also are distinct
in providing a detailed discussion of the origins of missingness for
159 predictors and exploring the intuition for why simple mean imputation
seems to perform well in ML applications, including using neural network
models and FHNW's regularized non-linear strategies.

The contemporaneous papers also differ in their predictor and stock
return datasets. In this respect, FHNW is the most similar to ours.
We both use 80+ predictors from the CZ open-source dataset and cover
all common stocks in CRSP. \citet{beckmeyer2022recovering} examine
143 characteristics from the \citet{jensen2021there} open source
dataset and restrict stock-months to those with data on at least 20\%
of the predictors. BLLP examine 45 characteristics from \citet{kozak2020shrinking}
and restrict stocks to those with at least one Compustat characteristic.
Given the size and breadth of these data, and the diversity of statistical
methods, the empirical results in these studies are complementary.

Our results provide a rigorous justification for the ubiquitous use
of mean imputation in machine learning papers. Almost every paper
that attempts to combine large sets of predictors imputes with cross-sectional
means or medians, including \citet*{gu2020empirical,freyberger2020dissecting,kozak2020shrinking}.\footnote{Other machine learning style papers that examine imputing with cross-sectional
averages include \citet{green2017characteristics,demiguel2020transaction,fulop2022option,han2021expected,azevedo2022stock,da2022statistical,simon2022deep,zhang2022dynamic,jensen2022machine}.} This result is important because the algorithms in these papers are
often complex, so a simple imputation algorithm is helpful for transparency.

Our PC analyses relate to the ongoing debate about the factor structure
of the cross-section of returns. Our findings are consistent with
the idea that there are many dimensions of predictability if one is
interested in all stocks (\citet*{green2014remarkable}; \citet*{bessembinder2021time})
and that a moderately strong factor structure is possible if one considers
only large stocks (\citet*{green2017characteristics}; \citet{lettau2020factors};
\citet{kozak2020shrinking}). Our findings complement \citet{lopez2020common},
who document large Sharpe ratios in strategies that hedge the principal
components of individual stock returns.

\section{The Structure and Origins of Missing Predictor Data\label{sec:sum}}

We describe the structure and origins of missingness among published
cross-sectional predictors. We also explain why imputation is required
for large scale machine learning studies.

\subsection{Cross-Sectional Predictor Data\label{sec:sum-data}}

Our predictors data begins with the August 2023 release of the CZ
dataset. This release contains 212 cross-sectional predictors published
in academic journals. We merge this data with CRSP and keep only common
stocks (SHRCD 10, 11, or 12) listed on the NYSE, NYSE MKT, or NASDAQ
(EXCHCD 1, 2, 3, or 31, 32, 33). This ensures that the missing data
problems we study apply to widely-studied common stocks, and do not
occur because of the rise of ETFs or other more exotic securities.
The screens we use follow \citet{banz1981relationship}, \citet{fama1992cross},
and \citet{bali2016empirical} among others.\footnote{Previous versions of this paper did not apply these screens. We thank
Jun Kyung Auh for pointing out this issue.}

We drop the 33 discrete predictors because the missing values in this
subset are closely related to the interpretation of these predictors
by CZ. Many of these predictors draw on event studies (e.g. exchange
switch), which CZ interpret as a binary firm characteristic (following
\citet{mclean2016does}). Others are based on double-sorts (e.g. analyst
recommendation and short interest). In both cases, CZ use missing
values for any stocks that belong in neither the long or short portfolio,
though there are other ways to interpret these studies.

We drop an additional 20 predictors because they have months with
fewer than 2 stock observations in the 1985-2021 sample. 18 of these
predictors use data that does not begin until after 1985. The remaining
predictors are junk stock momentum (Mom6mJunk) and firm age - momentum
(FirmAgeMom), which have small gaps in recent years. %
This screen ensures the predictor data is relatively well-behaved
and avoids the use of special techniques to handle predictor-months
with no data.

The final 159 predictors includes not only those based on CRSP and
Compustat, but also predictors that use IBES analyst forecast data.
This inclusion is important, as analyst forecasts may offer a distinct
dimension of predictability. Unfortunately, it excludes option-based
predictors, as option prices are not observed until the 1990s.

\subsection{Where Does Missingness Come From?\label{sec:sum-list}}

Table \ref{tab:short-list} presents a list of selected predictors
that illustrates the origins of missingness. The predictors are sorted
by the share of stocks with observations in 1985, a year that is fairly
representative of other years. The observed share ranges from 99.8\%
for size all the way down to 0.2\% for institutional ownership among
high short interest stocks. We curate this list from the full list
of 159 predictors, which can be found in the Internet Appendix (Table
IA.1).
\begin{center}
{[}Table \ref{tab:short-list} about here{]}
\par\end{center}

The most-observed predictors use only recent CRSP data. These predictors
include short-term reversal, idiosyncratic volatility, and size, and
are observed for nearly 100\% of stocks. Requiring a longer history
of CRSP data leads to a small but non-trivial share of missing values.
12-month momentum is observed for 92\% of stocks. A similar amount
of CRSP data is required for coskewness, beta, and two of the zero
trade predictors.

Requiring accounting variables drops the observed share into the 60\%
to 75\% range. Predictors that fall in this range include B/M and
asset growth. Many other accounting-based predictors fall in this
range, including employee growth, leverage, taxes, external financing,
and accruals.

Rates of observability below 50\% are driven by a wide variety of
issues. Most of these predictors require a long history of accounting
or CRSP data, such as return seasonality in years 11-15 or revenue
growth rank, which uses the past 5 years of sales growth. Long histories
of data are also required for intangible value, \citet*{titman2004capital}'s
investment predictor, and composite debt issuance.

But many of the less-observable predictors are missing values simply
because the data items they require are often missing. Predictors
that use analyst forecast data are observed for at most 50\% of stocks,
such as EPS forecast revision. Earnings surprise streak requires a
long history of EPS forecasts, leading to an even smaller share. Data
items are also often missing for less-common accounting variables,
like advertising expense.

Still other predictors have missing values due to a diverse range
of specialized filters. Asset tangibility applies only to manufacturing
firms, while IPO and age only applies to recent IPOs. Payout yield
drops stocks with non-positive payout. In CZ's interpretation, turnover
volatility drops stocks in the top two size quintiles, and operating
profitability drops stocks in the bottom size tercile. The original
Piotroski F-score (PS) study only examines stocks in the highest BM
quintile.

At the bottom of Table \ref{tab:short-list} are selected predictors
that are observed for less than 10\% of stocks in 1985. These very
low rates of observability are sometimes due to highly specialized
data. Conglomerate return uses Compustat segment data and customer
momentum uses the BEA input-output tables. Yet others combine specialized
data with the use missingness to substitute for interaction effects,
such as institutional ownership for high short interest.

As seen in these details, missingness is sometimes used as a substitute
for modeling non-monotonicity and interaction effects. Incorporating
these interactions is non-trivial. It requires going into the code
behind the individual predictors, removing the missingness assignments,
and carefully introducing new interaction variables. We leave this
work for future research.

\subsection{Summarizing Missingness for Individual Predictors\label{sec:sum-results}}

Table \ref{tab:miss-sum} describes missingness at the individual
predictor level. For each predictor, we compute the share of stocks
that have observed predictors in selected months between 1970 and
2000. Panel (a) shows order statistics across predictors.
\begin{center}
{[}Table \ref{tab:miss-sum} about here{]}
\par\end{center}

Though many cross-sectional predictability studies begin in 1963,
at least 5\% of our 159 predictors have no observations until about
1980. Several of these predictors draw on analyst forecast data, which
is not generally observed until 1985 (using IBES).

Missingness reaches a steady state in 1985, as the order statistics
are largely unchanged after this date. In this steady state, the typical
predictor is observed for 70\% of stocks, and 75\% of predictors are
observed for roughly 50\% of stocks. While this distribution suggests
that missing values are a relatively modest problem, we will see that
combining large sets of predictors compounds the issue.

This steady state suggests that missingness in the time series exhibits
structural breaks: predictor data is continually missing until the
data begins being published, after which the data is continually observed.

\subsection{Why Imputation is Often Required\label{sec:sum-setmiss}}

When combining predictors, one typically needs a numerical value for
all predictors for each stock-month in consideration. This requirement
compounds the magnitudes in Panel (a) of Table \ref{tab:miss-sum},
leading to a dramatically larger missingness problem.

This problem is illustrated in Figure \ref{fig:missmap}, which shows
the ``missingness map'' for the 159 predictors among common stocks
in June 1990. Roughly half of the map is non-shaded, indicating that
roughly half of the data is observed. This does not mean, however,
that roughly half of stocks can be used in a machine learning algorithm
without imputation. In general, only stocks with no missing values
can be used, corresponding to columns in the missingness map that
are entirely non-shaded. \emph{Zero} columns satisfy this requirement.
\begin{center}
{[}Figure \ref{fig:missmap} about here{]}
\par\end{center}

Panel (b) of Table \ref{tab:miss-sum} quantifies this problem. This
panel examines the prevalence of missing values when combining $J$
predictors, where $J$ ranges from 25 to 150. The $J=25$ row examines
the 25 predictors with the most observations in selected months. It
shows that one can combine 25 predictors with minimal missing data
issues. In June 1985, 96\% of predictor-stocks have observations and
86\% of stocks have predictor data for all of these 25 predictors.

But when combining for 75 or more predictors, one needs to take a
stand on imputation. As seen in the $J=75$ row, simply dropping stocks
with missing values would result in dropping all but 16\% of the data.
For $J=125$, essentially the entire dataset is gone if one follows
the traditional missing value handling used in individual predictor
studies.

Returning to Figure \ref{fig:missmap}, the complex missingness patterns
show that a general purpose missing data imputation is required. There
are dozens of distinct shapes in Figure \ref{fig:missmap}, suggesting
that dozens of equations are required to model the specific missingness
mechanisms. It is unlikely that an analyst could convincingly present
such a complicated model. The only alternative is to assume \citet{rubin1976inference}
ignorability, which leads to our focus on EM methods (Section \ref{sec:method-EM}).

\subsection{The Structure of Missingness\label{sec:sum-structure}}

Figure \ref{fig:missmap} sorts the predictors by the type of data
they focus on (according to the CZ documentation). This sort shows
that missingness occurs in blocks, organized by data type. Stocks
that are missing some accounting predictors are often missing every
single accounting predictor. The same pattern arises with analyst
forecast predictors, as seen in the dark vertical lines that span
the ``Analyst'' section of the plot.

Block structures also occur within data categories, consistent with
the fact that missingness often occurs due to the requirement of having
a certain amount of historical data. This block structure suggests
that extracting information about missing values from cross-sectional
observations may be difficult.

Missingness is extremely persistent. For predictor-stock combinations
that are missing in June 1990 (everything except for the lightest
shade), the vast majority were never observed before June 1990 (darkest
shade). Only a tiny share has observations in the past 12 months.
This structure implies that there is very little time-series information
about missing values.

This missingness structure leads us to focus on linear cross-sectional
imputations in our baseline methods. These methods focus on the part
of the data which contains more information about the missing values
(the cross-section), and they avoid over-parameterizing the model
when the amount of information in the cross-section is likely limited.

\section{Baseline Imputation Methods\label{sec:method}}

The missingness patterns imply that one \emph{must }impute data when
combining more than 100 predictors. This section describes our baseline
imputation methods: EM and simple mean.

These imputations are important benchmarks. EM is recommended in the
influential \citet{little2019statistical} textbook on missing data.
Simple mean imputation is the most common method in ML-style asset
pricing.

We examine several additional imputations in Section \ref{sec:broad}.

\subsection{Preprocessing\label{sec:method-boxcox}}

Before applying any imputation, we use the following preprocessing.
First, we winsorize across stocks symmetrically at the 1\% level for
each predictor-month. Second, we apply \citet{hawkins2017combining}'s
extension of \citet{box1964analysis} to make each predictor approximately
normal. Finally, we standardize predictors to have zero mean and unit
variance.

The Box-Cox transform is not commonly found in asset pricing, but
it is a natural extension of the log transformation that is typically
applied to size and B/M (\citet{fama1992cross}). More broadly, cross-sectional
predictability studies rarely impose a functional form on the relationship
between the predictor predictors and expected returns, so one might
as well standardize the predictors to be approximately normal.

\subsection{Cross-Sectional EM Imputation\label{sec:method-EM}}

There are $N$ stocks, and stock $i$ in month $t$ has a vector of
predictor values $X_{i,t}$. Missing values are represented by $X_{\text{\text{miss}}|i,t}$
and $X_{\text{\text{obs}}|i,t}$, which are the missing and observed
subvectors of $X_{i,t}$.\footnote{For example, if $X_{i,t}=\left[\begin{array}{cccc}
1 & 2 & 4 & 5\end{array}\right]^{\prime}$, and the first and third elements are missing, then $X_{\text{miss}|i,t}=[\begin{array}{cc}
1 & 3\end{array}]$ and $X_{\text{obs}|i,t}=[\begin{array}{cc}
2 & 4\end{array}]$. $X_{i+1,t}$ may have a different set of missing predictors, and
thus the length of $X_{\text{\text{miss}}|i+1,t}$ and $X_{\text{\text{obs}}|i+1,t}$
may be different compared to $X_{\text{\text{miss}}|i,t}$ and $X_{\text{\text{obs}}|i,t}$.}

Suppose $\hat{\Sigma}_{t}$ is a reasonable estimate for $\Sigma_{t}$,
the cross-stock covariance matrix of $X_{i,t}$. For example, one
could let $\hat{\Sigma}_{t}$ be the sample covariance across pairs
of stocks that have predictor data, as in BLLP. Then, an intuitive
way to impute $X_{\text{\text{miss}}|i,t}$ is to use:
\begin{align}
\hat{X}_{\text{\text{miss}}|i,t} & =\text{\ensuremath{\hat{\beta}_{i,t}^{\prime}}}X_{\text{\text{obs}}|i,t},\label{eq:em-regression}
\end{align}
where
\begin{align}
\hat{\beta}_{i,t} & \equiv\hat{\Sigma}_{\text{\text{obs}},\text{\text{obs}}|i,t}^{-1}\hat{\Sigma}_{\text{\text{obs}},\text{\text{miss}}|i,t},\label{eq:em-slope}
\end{align}
and $\hat{\Sigma}_{\text{\text{miss}},\text{\text{obs}}|i,t}$ and
$\hat{\Sigma}_{\text{\text{obs}},\text{\text{obs}}|i,t}$ are the
submatrices of $\hat{\Sigma}_{t}$ corresponding to the missing and
observed predictors for stock $i$. Analogous to the classic OLS formula
$\hat{\beta}=[X^{\prime}X]^{-1}X^{\prime}y$, Equations (\ref{eq:em-regression})-(\ref{eq:em-slope})
use the covariance between the LHS and RHS variables to forecast the
LHS. Equation (\ref{eq:em-regression}) omits the intercept since
we standardize variables to have mean zero (Section \ref{sec:method-boxcox}).

The problem with this approach is that it is not self-consistent.
That is, the imputed data would generally imply
\begin{align}
\hat{\Sigma}_{t} & \ne N^{-1}\sum_{i=1}^{N}\hat{X}_{i,t}^{\vphantom{}}\hat{X}_{i,t}^{\prime},\label{eq:em-inconsistency}
\end{align}
where $\hat{X}_{i,t}$ combines $\hat{X}_{\text{\text{miss}}|i,t}$
and $\hat{X}_{\text{\text{obs}}|i,t}$. In this case, it is unclear
whether you should be using $\hat{\Sigma}_{t}$ or $N^{-1}\sum_{i=1}^{N}\hat{X}_{i,t}^{\vphantom{}}\hat{X}_{i,t}^{\prime}$
to be imputing missing values.

A secondary issue is that Equations (\ref{eq:em-regression})-(\ref{eq:em-slope})
do not tell you how to deal with higher order missing terms. That
is, not only is $X_{\text{\text{miss}}|i,t}$ unknown, but so is $X_{\text{\text{miss}}|i,t}^{\vphantom{}}X_{\text{\text{miss}}|i,t}^{\prime}$.
The expected value of $X_{\text{\text{miss}}|i,t}^{\vphantom{}}X_{\text{\text{miss}}|i,t}^{\prime}$
is different than the outer product of Equation (\ref{eq:em-regression}),
and is potentially important for a rigorous estimate of $\Sigma_{t}$.
An intuitive way to estimate this higher order term is to use 
\begin{align}
\left[\widehat{X_{i,t}^{\vphantom{}}X_{i,t}^{\prime}}\right]_{\text{\text{miss}},\text{\text{miss}}|i,t} & =\Sigma_{\text{\text{miss}},\text{\text{miss}}|i,t}-\hat{\beta}_{i,t}^{\prime}\Sigma_{\text{\text{obs}},\text{\text{miss}}|i,t}+\hat{X}_{\text{\text{miss}}|i,t}^{\vphantom{}}\hat{X}_{\text{\text{miss}}|i,t}^{\prime},\label{eq:em-higher-order}
\end{align}
 which can be derived from Gaussian updating formulas.

The EM algorithm solves the self-consistency problem by iterating
between Equations (\ref{eq:em-regression})-(\ref{eq:em-slope}) and
Equation (\ref{eq:em-inconsistency}):
\begin{enumerate}
\item E-step: impute missing data with the observed data and the current
guess $\hat{\Sigma}_{t}$, using Equations (\ref{eq:em-regression})
and (\ref{eq:em-slope}).
\item M-step: estimate a new guess $\hat{\Sigma}_{t}^{\text{new}}$ by plugging
the imputed data from step 1 into the RHS of Equation (\ref{eq:em-inconsistency}).
Add to $\hat{\Sigma}_{t}^{\text{new}}$ the higher order correction
implied by Equation (\ref{eq:em-higher-order}) for submatricies corresponding
to $\hat{X}_{\text{\text{miss}}|i,t}^{\vphantom{}}\hat{X}_{\text{\text{miss}}|i,t}^{\prime}$.\footnote{Some descriptions of EM omit the higher order correction (e.g. \citet{roweis1997algorithms};
\citet{efron2016computer}).}
\end{enumerate}
These two steps are repeated until $\left\Vert \hat{\Sigma}_{t}^{\text{new}}-\hat{\Sigma}_{t}\right\Vert _{\infty}\le\delta$.
We use $\delta=10^{-4}$.

The E- and M-steps can be applied to almost any imputation and estimation
formulas. Our main results focus on the simple cross-sectional regression
formula (Equations (\ref{eq:em-regression})-(\ref{eq:em-slope})).
This method avoids look-ahead bias and is intuitive given the missingness
structure described in Section \ref{sec:sum-structure}. We examine
variations that assume a factor structure and take on time-series
information in Section \ref{sec:broad}.

More formally, this EM algorithm can be derived from maximum likelihood
(\citet{dempster1977maximum}), assuming that the process generating
the missing data is ``ignorable'' in the sense of (\citet{rubin1976inference}).
A sufficient condition for ignorability is that data is ``missing
at random,'' which is a technical condition that is best described
as the weakest general condition that leads to ignorability (\citet{little2021missing}).
This notion of ``missing at random'' is different than the one used
in \citet{xiong2022large}, whose notion of ``missing at random''
is often described as ``missing completely at random'' (\citet{little2019statistical}).
We provide a formal derivation of our EM algorithm and discuss these
technical issues in the Internet Appendix.

This EM algorithm is sometimes described as ``requiring'' the assumption
of normality (e.g. \citet{freyberger2021missing}, BLLP). But while
normality along with ignorability are \emph{sufficient} conditions
for the validity of EM, they are not necessary assumptions. Indeed,
simulation studies find this algorithm works well even if the data
are not normal (\citet{little1988robust,azen1989estimation,graham1999performance,king2001analyzing}).
This robustness is consistent with the intuition that EM is an iterative
form of OLS, and like OLS, is maximum likelihood under ideal conditions
but works well in real world settings. \citet{little1988robust} shows
that allowing for fat tails leads to the same algorithm but with smaller
weights on the extreme observations. As we winsorize our data before
imputation, such a fat tail extension should have little effect on
our results.

\subsection{Simple Mean Imputation\label{sec:method-simplemean}}

Simple mean imputation replaces missing values with the mean observed
value conditional on the predictor-month. Given our pre-processing
(Section \ref{sec:method-boxcox}), simple mean imputation amounts
to replacing missing predictor values with zeros, and median imputation
is equivalent.

Mean imputation could lead to significantly biased inferences (BLLP).
But Equations (\ref{eq:em-regression}) and (\ref{eq:em-slope}) show
this may not be the case. Mean imputation is the special case of EM
(and thus maximum likelihood) in which the off-diagonal terms of $\hat{\Sigma}_{t}$
are zero. In this case, $\hat{\Sigma}_{\text{miss},\text{obs}|i,t}$
is a matrix of zeros, the imputation slopes $\hat{\beta}_{i,t}$ is
a vector of zeros, and $\hat{X}_{\text{\text{miss}}|i,t}$ is also
a vector of zeros.

More generally, if the off-diagonal terms of $\Sigma_{t}$ are close
to zero for predictor pairs that can be used for imputation, then
mean imputation should be similar to EM and other algorithms focused
on $\Sigma_{t}$.

Studies of large sets of cross-sectional predictors find that the
off-diagonal terms of $\Sigma_{t}$ are mostly close to zero (\citet{green2017characteristics,ChenZimmermann2022}).
We verify this result in Section \ref{sec:result-impinfo}. Moreover,
the predictor pairs that can be used for imputation often cross data
categories (Figure \ref{fig:missmap}), implying the relevant covariances
are even smaller than those that have been documented. Lastly, assuming
the off-diagonal terms are small can be thought of as a structural
restriction, implied by the idea that predictors need to be novel
to be worthy of publishing (\citet{chen2020publication}).

Mean imputation also has the advantages of tractability and transparency.
These advantages are notable in our setting due to the number and
complexity of the missingness patterns in Figure \ref{fig:missmap}.
For each missingness pattern, EM imputation implies a different matrix
inversion, and every matrix inversion must be done in every iteration
of the algorithm. There are other general-purpose algorithms that
are more tractable, but it can be difficult to formally derive their
statistical properties under general assumptions (e.g. \citet{roweis1997algorithms}'s
pPCA). The transparency of mean imputation is also important, given
the complexity of the ML procedures that are applied to imputed data.

\section{EM vs Simple Mean Imputation\label{sec:result}}

This section provides the main results. We compare the performance
of EM vs mean imputation for predicting long-short returns in single
sorts (Section \ref{sec:result-onepredictor}) or PCR portfolios (Section
\ref{sec:result-pcr}). We explain why this performance is largely
similar (Section \ref{sec:result-impinfo}) and why EM can sometimes
underperform (Section \ref{sec:result-noise}).

We also show how \citet{huang2022scaled}'s scaled PCR leads to a
smaller dimensionality of expected returns, though imputations still
have little effect on inferences about expected returns (Section \ref{sec:result-spca}).

\subsection{Single-Predictor Strategies\label{sec:result-onepredictor}}

We start with single-predictor strategies because they are so familiar
to asset pricing researchers. It is unlikely that missing data handling
has a large effect here, but these results are helpful ``descriptive''
statistics before the main results.

Our single-predictor strategies go long the 500 stocks with the strongest
predictor and short the 500 stocks with the weakest predictor each
month. We do this for two missing data handling methods: (1) EM imputation
and (2) dropping stocks with missing values.

This design allows us to examine the most common method for handling
missing values (dropping missing) while simultaneously removing ``predictor
dilution'' effects that can occur when using imputed data. Intuitively,
the imputed data covers more stocks, and if these additional stocks
have more neutral predictors, the extreme deciles using imputed data
will display less predictability simply because of this additional
coverage. Fixing the number of stocks in each leg controls for this
issue.

Table \ref{tab:one-signal} shows the resulting distribution of mean
returns and Sharpe ratios. EM generally improves the distribution
of performance, but the typical effect is small, and it comes at the
cost of worse performance for many predictors. The distribution of
mean returns shifts to the right overall, but the left tail shifts
to the left, particularly in equal-weighted portfolios. Using Sharpe
ratios, the typical improvement is effectively zero, while both tails
shifts outward.
\noindent \begin{center}
{[}Table \ref{tab:one-signal} about here{]}
\par\end{center}

\subsection{159-Predictor Strategies Using PCR \label{sec:result-pcr}}

We now examine the more compelling setting for studying missing values:
strategies that combine information from many predictors. We combine
information using principal components regression (PCR), a simple
method for handling potentially correlated predictors and potential
overfitting problems.

Our PCR analysis forms ``real-time'' long-short strategies as follows:
For each month $t$ between January 1995 and December 2021, we apply
the following steps:
\begin{enumerate}
\item Separate stocks into three size groups (micro, small, and big) according
to market cap by month. Micro is below the 20th percentile NYSE market
cap and big is above the 50th percentile NYSE market cap, following
\citet{fama2008dissecting}.
\item For each size group, find the principal components (PCs) of the predictors
using data from the past 120 months. Then use OLS to predict stock
returns using the first $K$ PCs.
\item Form a portfolio that goes long the top decile of predicted returns
and short the bottom decile. Hold this position for one month.
\end{enumerate}
We make inferences on each size group separately because predictability
differs by market cap (\citet{fama2008dissecting}) and market liquidity
(\citet{chen2022zeroing}). Within each size group, OLS and PCA are
run pooling observations.\footnote{Pooling can be justified if $\left\{ N^{-1}\sum_{i=1}^{N}X_{i,t}X_{i,t}\right\} _{t=1,2,3,...}$
is weakly dependent, so that the pooled estimator converges to $E\left(X_{i,1}X_{i,1}^{\prime}\right)$.}

Figure \ref{fig:pca-bysize} shows the result. It plots the mean returns
(Panel (a)) and Sharpe ratios (Panel (b)) as a function of the number
of PCs used to forecast returns. The various lines show equal- or
value- weighting, and EM or simple mean imputation.
\noindent \begin{center}
{[}Figure \ref{fig:pca-bysize} about here{]}
\par\end{center}

Notably, the results depend relatively little on whether EM or mean
imputation is used. Echoing Table \ref{tab:one-signal}, EM leads
to high mean returns in equal-weighted portfolios, but the improvement
is minor to moderate and vanishes if many PCs are used. For value-weighted
portfolios, EM and simple mean imputation lead to very similar inferences
for most of the plot.

These results suggest that the simple mean imputation used in the
machine learning strategies of \citet{freyberger2020dissecting,gu2020empirical}
and others largely captures the potential returns and Sharpe ratios.
We verify this robustness for boosted tree and neural network strategies
in Section \ref{sec:broad}.

Figure \ref{fig:pca-bysize} also suggests an extremely high dimensionality
of expected returns. Mean returns and Sharpe ratios increase in the
number of PCs, up to at least 40 PCs. This dimensionality may be overstated,
however, as PCA does not incorporate return information, and can be
thought of as an ``unsupervised learning'' model. We revisit this
issue in Section \ref{sec:result-spca}, where we examine a ``supervised''
version of PCA.

\subsection{Imputation Information and Predictor Correlations\label{sec:result-impinfo}}

Why do EM and simple mean imputation lead to similar inferences about
expected returns?

The answer is that the observed predictors contain little information
about the missing predictors. This fact is illustrated in Figure \ref{fig:cor},
which shows the distribution of correlations between pairs of predictors.
The ``Observed'' lines use all pairs of stocks that have predictor
data in the given month (``available case'' or ``pairwise complete''
correlations). The ``EM Algo'' lines use the correlations estimated
from EM.
\noindent \begin{center}
{[}Figure \ref{fig:cor} about here{]}
\par\end{center}

The correlations cluster near zero, consistent with \citet{ChenZimmermann2022}.
As a result, mean imputation should be a reasonable approximation
of EM (Section \ref{sec:method-simplemean}). Intuitively, the low
correlations imply that the observed value of one predictor tells
you relatively little about a missing predictor. Similarly, principal
components are relatively uninformative. 10 PCs capture only 40\%
of total variance.\footnote{The relationship between PCA and imputation slopes is difficult to
formalize but we show high dimensionality is associated with small
imputation slopes in simulations in the Internet Appendix.} These results are stable over time and do not depend on whether you
use available case or EM correlations.\footnote{Though the observed and EM distributions look identical, we show in
the Internet Appendix that they differ slightly.} Consistent with the interpretation that the observed data is uninformative,
we find that the imputation slopes (Equation (\ref{eq:em-slope}))
cluster close to zero (Internet Appendix Figure IA.4).

Figure \ref{fig:cor} actually overstates the potential information
gains. As discussed in Section \ref{sec:sum-setmiss}, missingness
tends to occur in blocks organized around the data source. If a stock
is missing an accounting-based predictor, it is likely to be missing
all accounting-based predictors. So while earnings-to-price and book-to-market
may be correlated, this correlation typically cannot be used to impute
book-to-market. Instead, one would need to impute with a more distant
predictor like coskewness or trading volume, which are unlikely to
tell us much about book-to-market.

Similarly, while Figure \ref{fig:cor} focuses on cross-sectional
information, one might argue that there is time-series information
from, say, the persistence in book-to-market. However, missingness
also tends to occur in blocks organized around time (Figure \ref{fig:missmap}).
If book-to-market is missing now, it is extremely likely to be missing
last year, so this time-series correlation is also not informative.

\subsection{Imputation Noise and Pitfalls\label{sec:result-noise}}

EM imputation not only adds little information, it also introduces
noise. This noise can lead to poor forecasting performance if the
data is not carefully handled.

Figure \ref{fig:impute-err} illustrates the noise introduced by imputation.
It shows imputation errors from a masking exercise: we randomly mask
10\% of observed stock-predictor values, impute with EM, and calculate
the error as the difference between the masked value and the imputed
value.
\noindent \begin{center}
{[}Figure \ref{fig:impute-err} about here{]}
\par\end{center}

For stocks with below-median market cap, the RMSE is at least 0.70.
For comparison, imputing with simple means would lead to an RMSE of
1.00, since the data is standardized (Section \ref{sec:method-boxcox}).
Thus, the imputation errors are quite large, especially among small
stocks.

Indeed, the RMSEs in Figure \ref{fig:impute-err} are surely understated,
as our random masking does not account for the fact that accounting
predictors are typically missing together. To mask, we draw all observed
predictor-stock combinations with equal probability, so we will often
mask book-to-market without masking the other accounting-based predictors.
Thus, the imputation exercise in Figure \ref{fig:impute-err} can
``cheat,'' by imputing book-to-market with earnings-to-price, even
though in the real world these variables are almost always missing
together.

Imputation noise can lead to poor forecasting performance, as seen
in Figure \ref{fig:pca-pooled}. This figure repeats the long-short
portfolio exercise from Section \ref{sec:result-pcr}, but instead
of estimating predicted returns by size group, it runs pooled estimations,
estimating a single model for all size groups simultaneously. Thus,
it ignores the fact that predictability is stronger in small stocks,
as well as the fact that imputation errors are larger for small stocks.
\noindent \begin{center}
{[}Figure \ref{fig:pca-pooled} about here{]}
\par\end{center}

In value-weighted portfolios, EM leads to much worse performance than
simple mean imputation if more than 75 PCs are used. Importantly,
simple mean imputation leads to value-weighted returns as high as
25\% per year, while EM imputation tops out at 20\%. This problem
is much smaller in equal-weighted portfolios, in which returns top
out at around 53\% per year, regardless of the imputation method.

This underperformance is consistent with the pattern in imputation
errors. EM underperformance happens when many PCs are used, forecasts
are pooled, and performance is evaluated using value-weighting. In
this setting, the large imputation errors in small stocks affect the
forecasts of large stocks, and these large-stock forcasts are the
focus of the performance evaluation.

These results illustrate a pitfall of sophisticated imputations. The
noise introduced by sophisticated imputation may outweigh the information
gained. This risk is higher if forecasts are not aligned with performance
evaluation, as in Figure \ref{fig:pca-pooled}'s value-weighted results.
Section \ref{sec:broad} shows that this pitfall is not just a property
of EM imputation and PCR, but is also seen using other methods.

This pitfall leads us to recommend simple mean imputation for cross-sectional
asset pricing research. While sophisticated imputations more fully
incorporate information, they also introduce noise and can lead to
underperformance. Given the complexity of most imputation and ML forecasting
methods, our judgment is that the benefits of simple mean imputation
outweigh the costs.

\subsection{\citet{huang2022scaled}'s Scaled PCR and the Dimensionality of Expected
Returns\label{sec:result-spca}}

The PCR results (Figures \ref{fig:pca-bysize} and \ref{fig:pca-pooled})
imply that there are dozens of PCs that contribute to the cross-section
of expected returns. This section shows that the dimensionality of
expected returns is actually subtle and depends on multiple measurement
decisions. To show this result, we repeat the portfolio formation
exercise in Section \ref{sec:result-pcr}, but use \citet{huang2022scaled}'s
scaled PCA instead of standard PCA.

We apply \citet{huang2022scaled}'s scaled PCA as follows: For each
predictor $j$, we estimate $r_{i,t}=\alpha_{j}+\gamma_{j}X_{i,j,t-1}+\varepsilon_{i,j,t}$
using OLS, where $X_{i,j,t-1}$ is the value of predictor $j$ for
stock $i$ in month $t-1$. Then we define the scaled predictor $\dot{X}_{i,j,t}\equiv\hat{\gamma}_{j}X_{i,j,t}$
and apply standard PCA to $\dot{X}_{i,j,t}$ to obtained scaled PCs.
This rescaling provides a simple and intuitive method for incorporating
return information into the PC estimation. \citet{huang2022scaled}
show that scaled PCA outperforms standard PCA using theory, simulation,
and a macroeconomic forecasting exercise.

The performance of scaled-PCR portfolios is shown in Figure \ref{fig:spca-bysize}.
Compared to Figure \ref{fig:pca-bysize}, the implied dimensionality
is much smaller, especially if it is applied to EM-imputed data. In
this specification, only 30 PCs are required to capture the potential
equal-weighted returns, compared to about 70 PCs using standard PCR.
Focusing on very large stocks implies a much smaller dimensionality.
Scaled PCR requires only about 15 PCs to capture the potential value-weighted
returns (Panel (b)), compared to about 30 PCs using standard PCR.
\noindent \begin{center}
{[}Figure \ref{fig:spca-bysize} about here{]}
\par\end{center}

Still, mean returns depend relatively little on how missing data is
handled. Through most of Figure \ref{fig:spca-bysize}, the lines
for EM and simple mean imputation largely overlap. And the potential
``real-time'' mean returns top out at around 50\% per year equal-weighted
or 20\% value-weighted. Whether the underlying data is imputed with
EM or just simple means does not matter. We find broadly similar results
using CAPM and FF6 alphas in the Appendix (Figure \ref{fig:scpa-alpha}).

An important exception to this invariance is in value-weighted portfolios
that use more than 125 PCs. Here, EM imputation underperforms, consistent
with the idea that the estimation noise using EM can exceed the information
gained (Section \ref{sec:result-noise}).

\section{Comparing Six Imputations using Six Return Forecasting Methods\label{sec:broad}}

Our baseline results examine just two imputation methods and two return
forecasting methods. This section expands the analysis in both directions.

We examine four additional imputations, including ones that incorporate
time-series information. We examine four additional return forecasting
methods, including two based on neural networks.

The main results continue to hold: ad-hoc imputations and more sophisticated
imputations lead to largely similar inferences about expected returns.

\subsection{Four Additional Imputation Methods\label{sec:broad-impmethod}}

We examine the following additional imputations:
\begin{enumerate}
\item \textbf{Cross-sectional EM on AR1 Residuals: }We estimate an AR1 model
on Box-Cox transformed predictors using the past 60 months of data
and apply cross-sectional EM to the residuals. To impute missing predictors,
we add the AR1 prediction to the cross-sectional EM-imputed residuals.
This algorithm takes on time-series information, which BLLP find is
important in their imputation error tests.
\item \textbf{Practical EM for Probabilistic PCA: }We assume Box-Cox transformed
predictors follow a 10-dimensional factor structure and impute using
\citet{roweis1997algorithms}'s ``practical'' EM algorithm for missing
data. This practical EM algorithm simply imputes missing predictors
using the factor model prediction in the E-step, and as such is not
a maximum likelihood estimate (it ignores higher order terms like
Equation (\ref{eq:em-higher-order})). However, it is easily implemented
using the pcaMethods R library and provides a regularized alternative
to our cross-sectional EM imputation.
\item \textbf{Industry - Size Decile Means:} Within each 3-digit SIC code,
we sort stocks into size deciles. We then replace missing Box-Cox
transformed predictors with their means within each industry-size
group. This imputation applies the intuition that industry and size
are important determinants of a firm's economics.
\item \textbf{Last Observed: }Replace missing values with the last observed
value within the last 12 months. If there is no observed value within
the last 12 months, we replace missing values with cross-sectional
means. This imputation takes on time-series information while avoiding
using stale information. It is also easy to implement
\end{enumerate}
For additional details see Appendix \ref{sec:app-add-imp}.

\subsection{Four Additional Return Forecasting Methods\label{sec:broad-foremethod}}

We examine the following additional return forecasting methods:
\begin{enumerate}
\item \textbf{Simple OLS: }We simply regress future stock returns on predictors
using OLS without any regularization. Simple OLS provides a helpful
benchmark.
\item \textbf{Gradient-Boosted Regression Trees (GBRT): }We model future
stock returns using as the average of $B$ shallow regression trees.
The model is fitted by gradient boosting: starting from an existing
tree, the tree is ``boosted'' by adding a new tree, where the new
tree is chosen to step along the gradient of a squared-error loss
function, and this process is continued $B$ times. $B$ and other
hyperparameters are tuned following \citet{gu2020empirical} and \citet{li2023realtime},
and the algorithm is implemented with Microsoft's open source lightGBM
library.
\item \textbf{3-Layer Neural Network (NN3): }We model the relationship between
future stock returns and predictors as the average of an ensemble
of feed-forward neural networks with three hidden layers. We fit the
model using an Adam optimizer with batch normalization and early stopping.
Tuning and other specification details follow \citet{gu2020empirical},
who find that neural network performance peaks at three hidden layers.
We implement the algorithm with Google's open source Tensorflow and
Keras libraries.
\item \textbf{1-Layer Neural Network (NN1): }This forecast uses the same
model as NN3, but with only one hidden layer. This model highlights
the robustness of NN3, which we find is by far the best-performing
forecasting method.
\end{enumerate}
We apply these methods either to all stocks at the same time or by
size group. For additional details on the neural network specifications
see Appendix \ref{sec:app-add-fore}.

\subsection{Out-of-Sample Forecasting and Hyperparameter Tuning\label{sec:broad-oostune}}

We impute missing values using one of the six methods and then construct
``out-of-sample'' portfolios using one of the six forecasting methods.

Portfolio construction uses the following procedure. For each June
$t$ between 1995-2021, we
\begin{enumerate}
\item Separate the data between January 1985 and $t$ into a training sample
and a validation sample. The validation sample is either the latter
half of the data or the past 12 years, whichever sample is shorter.
The remainder of the data is the training sample.
\item For each size group (micro, small, big), tune hyperparameters and
fit the the return model.
\begin{enumerate}
\item For all hyperparameter sets in a pre-selected set-of-sets, fit returns
to lagged imputed data using the training sample and measure the RMSE
in the validation sample.
\item Using the hyperparameter set with the smallest RMSE, fit returns to
lagged imputed data from January 1985 to $t$.
\end{enumerate}
\item For each month $t,t+1,...,t+11$, sort stocks on the predicted return
and form long-short decile portfolios.
\end{enumerate}
This procedure follows \citet{gu2020empirical} closely, though we
add the size group separation, which we find generally improves performance
(see Section \ref{sec:result-noise}). We also differ in requiring
that at least half of the January 1985 to $t$ subsamples are used
for training. Gu et al. do not need this requirement because their
data goes back to 1957.

With the exception of OLS, the forecasting methods require hyperparameter
tuning. We tune hyperparameters following \citet{gu2020empirical}
as closely as possible (see their Appendix Table A.5). For PCR and
sPCR, we select the number of PCs from $\left\{ 10,30,50,70,90\right\} $.
For GBRT, the depth is 1 or 2, the learning rate is 0.01 or 0.10,
and the number of trees is in $\{1,250,500,750,1000\}$. For NN1 and
NN3, the L1 penalty is either $10^{-5}$ or $10^{-3}$, the learning
rate is 0.001 or 0.010, and other hyperparameters are fixed: batch
size is 10,000, epochs is 100, early stopping patience is 5, the ensemble
number is 10, and we use the Adam optimizer with default parameters
in the Keras interface to TensorFlow.\footnote{We differ from \citet{gu2020empirical} by using the standard L2 objective
in GBRT rather than a Huber loss. Our implementation of GBRT uses
lightGBM, which currently has poor documentation of the objective
function, and we find that objective = 'huber' works poorly on simulated
and empirical data. \citet{li2023realtime} does not mention using
Huber loss.}

\subsection{Results Across a Broad Array of Methods}

Table \ref{tab:imp-fore-grid} shows the result across all combinations
of the 6 imputations (rows) and 6 forecasting methods (columns). For
each forecasting method, we examine estimating separately for micro,
small, and big stocks (``By Size'') or estimating all stocks together
(``Pool''). We examine both equal-weighted (Panel (a)) and value-weighted
(Panel (b)) returns.
\noindent \begin{center}
{[}Table \ref{tab:imp-fore-grid} about here{]}
\par\end{center}

\subsubsection{Performance Across Imputations}

The central feature of Table \ref{tab:imp-fore-grid} is that returns
depend little on the imputation method. Neural networks outperform
other forecasting methods (columns) and returns are much higher in
equal-weighted portfolios (Panel (a)). But moving across imputations
(rows), there is little change in mean returns. For example, applying
a one-layer neural network (NN1) to mean imputation leads to equal-weighted
returns of 65\% per year, compared to 64\% imputing with EM on AR1
residuals. Similarly, applying a one-layer neural network to mean
imputation leads to value-weighted returns of 40\% per year, compared
to 39\% per year using EM on AR1 residuals. This strong performance
for mean imputation is consistent with the idea that the observed
data provide little information about the missing data (\ref{sec:result-impinfo}).

Mean imputation \emph{out}performs in some settings. This outperformance
is concentrated in value-weighted portfolios using pooled return forecasts
(Panel (b), ``Pool'' columns). This result is consistent with the
idea that sophisticated imputations can lead to additional noise in
small-cap stocks, which can lead to poor forecasting performance if
the data is not carefully handled (Section \ref{sec:result-noise}).
This pitfall leads us to recommend simple mean imputation for cross-sectional
asset pricing research.

There is one setting in which mean imputation can lead to significant
underperformance. Using GBRT, mean imputation can lead to noticeably
lower returns compared to EM, perhaps because regression trees deal
poorly with the degenerate distributions that result from mean imputation.
One should not over-interpret this result, however, since tree-based
algorithms often have their own specialized methods for handling missing
values, which we do not employ because of our goal of evaluating ``general-purpose''
imputations.\footnote{lightGBM can handle missing values by adding a second greedy stage
to each split assessment.}

Adding time-series information to the imputations has little effect.
The ``Last Obs'' row differs little from the other ad-hoc imputations
and similarly the ``EM AR1'' row differs little from the other EM
imputations. This result is consistent with the fact that missingness
occurs in large time-series blocks (Figure \ref{fig:missmap}). It
also illustrates the robustness of our main results to the incorporation
of time-series information, which BLLP find is important for imputation
accuracy under simulated missingness patterns. We also find that the
BLLP local backward cross-sectional imputation has little effect on
estimated expected returns (Appendix Figure \ref{fig:bllp}).

Similarly, regularizing EM by using pPCA has little effect. The ``pPCA10''
rows differ little from the other EM rows. We also find that using
industry / size means has little effect.

\subsubsection{Improvements due to Forecasting by Size Group}

A second theme in Table \ref{tab:imp-fore-grid} is that forecasting
by size group significantly improves performance in value-weighted
portfolios. Using PCR or sPCR, value-weighted mean returns increase
from around 20\% per year, to around 33\% per year when forecasting
by size group. Non-trivial gains from forecasting by size group are
also seen in OLS- or GBRT-based value-weighted portfolios.

These results are consistent with the fact that predictability differs
by market cap and liquidity (\citet{fama2008dissecting}; \citet{chen2022zeroing}).
We also find that the dimensionality of expected returns differs significantly
across size groups (Section \ref{sec:result-pcr}). Thus, the forecasting
model should vary by size group, and assuming that the expected return
structure found among big stocks is the same as those found in small
stocks will lead to sub-optimal forecasts.

Indeed, we find that our hyperparameter tuning, which is performed
by size group (Section \ref{sec:broad-foremethod}), finds a much
lower dimensionality for large stocks. This variable dimensionality
allows the PCR and sPCR value-weighted returns in Table \ref{tab:imp-fore-grid}
to far exceed the value-weighted returns in Figures \ref{fig:pca-bysize}
and \ref{fig:spca-bysize}, which fix the dimensionality for across
size groups.%

\subsubsection{Performance Across Forecasting Methods}

A third theme from Table \ref{tab:imp-fore-grid} is that ML-style
methods can \emph{under}perform relative to simple OLS. Moving across
the columns of Table \ref{tab:imp-fore-grid}, outperformance relative
to OLS is only seen using neural network forecasts.

These results may be surprising, given that many papers tout the outperformance
of machine learning methods (\citet{gu2020empirical,freyberger2020dissecting,simon2022deep}).
However, they are consistent with these papers once one examines the
details. \citet{gu2020empirical} find that applying PCR and GBRT
to 94 predictors leads to returns that are not much larger than those
obtained from applying linear regression to just three predictors.
\citet{freyberger2020dissecting} and \citet{simon2022deep} both
examine continuous non-linear models of returns, which are differ
substantially from linear PCR models and the discontinuous GBRT models.

Moreover, the underperformance of PC-based methods is intuitive given
the fact that mean returns are largely monotonic in the number of
PCs used (Figures \ref{fig:pca-bysize}, \ref{fig:pca-pooled}, and
\ref{fig:spca-bysize}). OLS always uses all PCs in the data. In contrast,
PC-based methods need to tune the number of PCs, which can lead to
too few PCs being used.

The outperformance of neural network forecasts is large. Mean returns
increase from around 50 percent per year to 65 percent using equal-weighting,
and from around 30 percent per year to 40 percent using value-weighting.
In fact, we find that other neural network architectures work quite
well also, and that performance is fairly robust to ensembling and
other tuning choices if early stopping is set following \citet{gu2020empirical}.
These results support the strong performance of neural networks found
in other papers (\citet{gu2020empirical,simon2022deep}) and shows
that this strong performance is robust to imputation methods.

\section{Conclusion}

In the age of machine learning, the standard practice of dropping
stock-months with missing predictors is often untenable. Previous
ML studies apply ad-hoc imputations or data adjustments, with little
discussion of their motivation or study of alternatives. The goal
of our paper is to provide guidance on this problem.

We recommend using simple mean imputation for ML studies. This recommendation
may seem surprising, since mean imputation essentially ignores all
information about the missing data. However, the missingness and the
covariance structures in cross-sectional predictor data imply that
there is little information about missing values. As a result, one
might as well ignore all information. Consistent with this idea, we
find mean imputation leads to similar portfolio returns and Sharpe
ratios across a broad array of imputation and return forecasting methods.

Our PC analyses relate to the ongoing debate about the factor structure
of the cross-section of returns. Our findings are consistent with
a very high dimensionality for all stocks. In contrast, there is the
potential for a strong factor structure among very large stocks. We
also find that the results depend on the methodology used to compute
PCs, with supervised methods like \citet{huang2022scaled} implying
far fewer PCs are required to fully capture the cross-section.

\pagebreak{}

\pagebreak{}

\appendix

\section{Appendix\label{sec:app}}

\setcounter{table}{0} \renewcommand{\thetable}{A.\arabic{table}}
\setcounter{figure}{0} \renewcommand{\thefigure}{A.\arabic{figure}}

\subsection{Additional Imputation Method Details\label{sec:app-add-imp}}

\subsubsection{Cross-Sectional EM with Time-Series Information\label{sec:app-add-imp-emar1}}

Our main EM imputation handles each cross-section independently. This
imputation incorporates time-series information.

We impute in three steps for each ``forecasting month'' $\tau$:
\begin{enumerate}
\item Using OLS estimate $\hat{\phi}_{j}$ for the AR1 model
\begin{align}
X_{i,j,t} & =\phi_{j}X_{i,j,t-h}+\varepsilon_{i,j,t}\label{eq:ts-x-ar1}
\end{align}
where $X_{i,j,t}$ is Box-Cox transformed predictor (as in Section
\ref{sec:method-simplemean}), $h\in\left\{ 1,3,12\right\} $ is the
periodicity of the underlying data updates (based on \citet{ChenZimmermann2022}'s
documentation), $\phi_{j}$ is the persistence of predictor $j$,
and $\epsilon_{i,j,t}$ is a zero mean shock. In this step, we use
observed data from month $\tau-59$ to $\tau$.
\item Using EM, estimate $\hat{\Omega}_{\tau}$ for 
\begin{align}
\hat{\varepsilon}_{i,\tau} & \equiv\left[\begin{array}{c}
\hat{\varepsilon}_{i,1,\tau}\\
\hat{\varepsilon}_{i,2,\tau}\\
...\\
\hat{\varepsilon}_{i,J,\tau}
\end{array}\right]\sim\text{MVN}\left(0,\Omega_{\tau}\right).\label{eq:ts-ep-mvn}
\end{align}
where $\hat{\varepsilon}_{i,j,\tau}\equiv X_{i,j,\tau}-\hat{\phi}_{j}X_{i,j,t-h}$.
Assuming standard regularity conditions, this two-step estimation
is consistent (\citet{wooldridge1994estimation}).
\item For missing $X_{i,j,\tau}$, impute by combining the prediction of
the AR1 model with the residual estimate---that is, use $\hat{\phi}_{j}X_{i,j,t-h}+\hat{\varepsilon}_{i,j,\tau}$
(if $X_{i,j,t-h}$ is observed) or $\hat{\varepsilon}_{i,j,\tau}$
(if $X_{i,j,t-h}$ is not observed).
\end{enumerate}

\subsubsection{Practical EM for Probabilistic PCA}

We did not restrict the covariance matrix in our main EM imputation.
This imputation imposes a factor structure and estimates using a variant
of EM (\citet{roweis1997algorithms}, see also \citet{tipping1999probabilistic}
).

We estimate the following model:
\begin{align}
X_{i,t} & =\Lambda_{t}F_{i,t}+\varepsilon_{i,t}\label{eq:ppca-1}\\
F_{i,t} & \sim\text{MVN}\left(0,I\right)\nonumber \\
\varepsilon_{i,t} & \sim\text{MVN}\left(0,\sigma^{2}I\right)\label{eq:ppca-2}
\end{align}
where $X_{i,t}$ is a $J$-dimensional vector of predictors for stock
$i$ in month $t$, $F_{i,t}$ is a $K$-dimensional vector of factor
realizations for stock $i$ in month $t$, $\Lambda_{t}$ is a $J\times K$
matrix that maps factors into predictors, and $\sigma^{2}$ is a scalar.

The estimation alternates between two steps for each forecasting month
$t$:
\begin{enumerate}
\item E-step: Given a guess for $\Lambda_{t}$ and $F_{i,t}$, replace the
missing components of $X_{i,t}$ with $\Lambda_{t}F_{i,t}$. Find
the expectations of $F_{i,t}$ and $F_{i,t}F_{i,t}'$ conditional
on $\Lambda_{t}$, $\sigma$, and $X_{i,t}$.
\item M-step: Plug the expectations of $F_{i,t}$ and $F_{i,t}F_{i,t}'$
into the log-likelihood assuming that all $X_{i,t}$ and $F_{i,t}$
are observed, and maximize to obtain new $\Lambda_{t}$ and $F_{i,t}$.
\end{enumerate}
The algorithm iterates these two steps until the objective in the
M-step no longer improves.

As in the case of our baseline model (Equation (\ref{eq:em-slope})),
Equations (\ref{eq:ppca-1})-(\ref{eq:ppca-2}) imply that the ignorable
quasi-likelihood can be optimized using EM (\citet{tipping1999probabilistic})
(see Internet Appendix A.2). This algorithm does not follow the full
EM method, as it does not include higher moments of missing $X_{i,t}$
in the E-step. However, the computer science literature finds that
less formal variants of EM work well on this model using simulated
data (\citet{roweis1997algorithms}; \citet{ilin2010practical}).
Our implementation uses the R library pcaMethods, which follows \citet{roweis1997algorithms}'s
algorithm.

\subsubsection{\citet*{bryzgalova2022missing}'s Factor Model with Added Time-Series
Information}

BLLP's local backward cross-sectional model is similar to the imputations
we study in that it can be used to form forecasts without look-ahead
bias. They emphasize the importance of time-series information, so
we examine how their method for adding time-series information affects
our scaled-PCA results here. We find similar results in our other
forecasting methods but show just the scaled-PCA results for brevity.

\begin{figure}[h!]
\caption{\textbf{BLLP's Imputation and Scaled-PCA Returns}}
\label{fig:bllp} \pdfbookmark{Figure A.5}{bllp} Each month, we forecast
returns using scaled principal components regression and form long-short
deciles portfolios. `BLLP loc B-XS' imputes using the local backward
cross-sectional model in BLLP, where we use six factors to be consistent
with Section 4.1 of BLLP, and lags used in the second stage regression
are restricted to relevant predictor update months based on documentation
for the \citet{ChenZimmermann2022} dataset.\textbf{ Interpretation:
}Taking on time-series information following BLLP results in similar
estimates as simple mean imputation.

\vspace{0.15in}

\centering 
\subfloat[Mean Returns]{\includegraphics[width=0.9\textwidth]{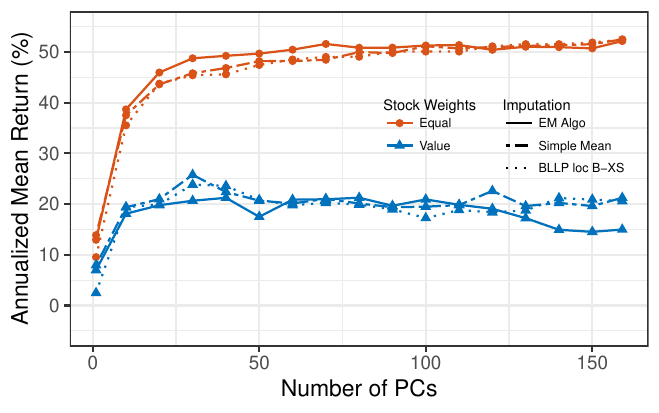}}\\
\subfloat[Sharpe Ratios]{\includegraphics[width=0.9\textwidth]{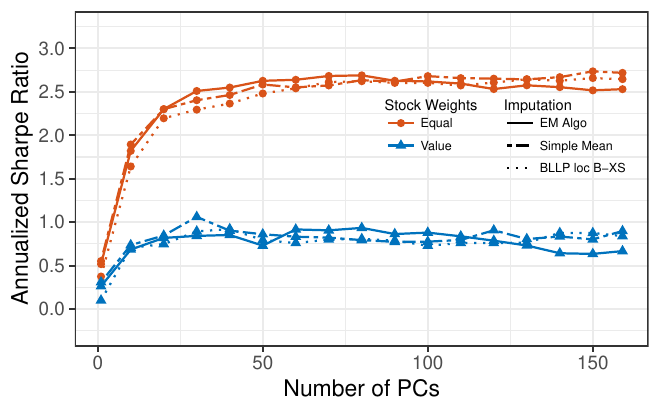}}   
\end{figure}

Our implementation of BLLP's local backward cross-sectional imputation
proceeds as follows. For each month of predictor data:
\begin{enumerate}
\item Compute the available case cross-sectional covariance matrix of predictors.
If there are no overlapping observations for a pair of predictors,
we use zero as the covariance matrix entry.
\item Form a matrix using the $K$ eigenvectors with the largest eigenvalues.
The rows of these matrix are the ``loadings'' of each predictor.
\item Regress predictor-stock values on the predictor loadings. The slopes
are called ``characteristic factors'' and there should be $K$ for
each stock.
\item Regress predictors on characteristic factors and lagged predictors
that are observed within the last 24 months. We only used lagged predictors
that have signal updates according to the Chen-Zimmermann documentation
(only use lags 12 and 24 for annual predictors).
\item Use the slopes from step 4 to impute missing values.
\end{enumerate}
We apply scaled-PCA (by size) and form portfolios. The results are
in Figure \ref{fig:bllp}.
\begin{center}
{[}Figure \ref{fig:bllp} about here{]}
\par\end{center}

\subsection{Additional Forecasting Method Details\label{sec:app-add-fore}}

GBRT details are fully described in Sections \ref{sec:broad-foremethod}
and \ref{sec:broad-oostune}.

Neural network details follow \citet{gu2020empirical}. NN1 has a
single hidden layer of 32 neurons and NN3 has hidden layers with 32,
16, and 8 neurons. All layers are connected with the ReLU activiation
function. The objective is L2 with a L1 penalty to weight parameters
and is minimized using the Adam extension of Stochastic Gradient Descent
under early stopping with a patience parameter of 5 in batches of
10,000, for 100 epochs. The L1 penalty and learning rate are tuned
as described in the main text. Last, we form final return forecasts
as the ensemble average of 10 neural network forecasts.

\subsection{Appendix Exhibits\label{sec:app-add-results}}

\clearpage\pagebreak{}

\clearpage\pagebreak{}

\clearpage\pagebreak{}
\begin{figure}[h!]
\caption{\textbf{Missing Data Effects on Alphas from Scaled PCA}}
\label{fig:scpa-alpha} \pdfbookmark{Figure A.6}{alpha} Each month,
we forecast returns using scaled principal components regression and
form long-short deciles portfolios, as in Figure \ref{fig:spca-bysize}.
We then measure CAPM alphas (Panel (a)) or Fama-French 5 factor +
momentum alphas (Panel(b)).\textbf{ Interpretation: }Simple mean imputation
also performs well when examining alphas.

\vspace{0.15in}

\centering 
\subfloat[CAPM]{\includegraphics[width=0.9\textwidth]{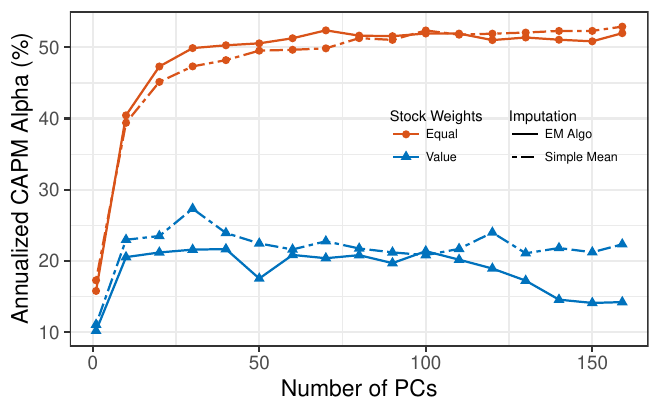}}\\
\subfloat[FF6]{\includegraphics[width=0.9\textwidth]{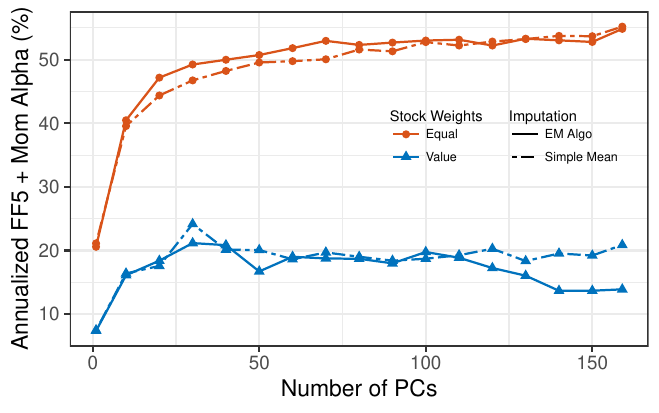}}   
\end{figure}

\clearpage\pagebreak{}

\bibliographystyle{elsarticle-harv}
\bibliography{0C__Dropbox_Missing_Data_arxiv-sub-2024-01_paper-missing-signals_missing-signals}

\pagebreak{}

\setcounter{table}{0} \renewcommand{\thetable}{\arabic{table}}
\setcounter{figure}{0} \renewcommand{\thefigure}{\arabic{figure}}

\clearpage\pagebreak{}
\begin{table}
\caption{\textbf{Missingness in 1985 for Selected Predictors}}
\label{tab:short-list} \pdfbookmark{Table 1}{short-list}

\begin{singlespace}
\noindent Predictors are selected to illustrate the causes of missingness
(see discussion in text). `\% Obs' is the mean share of stock-months
for which the predictor is observed in 1985, for common stocks listed
on the NYSE, NYSE MKT, and NASDAQ. The full list of 159 continuous
predictors with at least 2 observations in every month between 1985
and 2021 is in the Internet Appendix (Table IA.1). \textbf{Interpretation:}
Missingness is due to the availability of the underlying data (accounting,
analyst forecast, etc), the requirement of having a long history of
underlying data (Ret Seasonality Yr 11-15), and is also imposed to
approximate interaction effects (IPO and Age).
\end{singlespace}
\begin{centering}
\vspace{0ex}
\par\end{centering}
\centering{}\setlength{\tabcolsep}{1ex} \small
\begin{center}
\begin{tabular}{lllr}
  \toprule
Predictor & Reference & Data Focus & \% Obs \\ 
  \midrule
Size & Banz 1981 & Price & 99.8 \\ 
  12-Month Momentum & Jegadeesh and Titman 1993 & Price & 91.6 \\ 
  Book-to-Market & Stattman 1980 & Accounting & 74.4 \\ 
  Asset Growth & Cooper, Gulen and Schill 2008 & Accounting & 67.1 \\ 
  EPS Forecast Revision & Hawkins, Chamberlin, Daniel 1984 & Analyst & 46.0 \\ 
  Ret Seasonality Yr 11-15 & Heston and Sadka 2008 & Price & 45.3 \\ 
  Revenue Growth Rank & Lakonishok, Shleifer, Vishny 1994 & Accounting & 40.5 \\ 
  Payout Yield & Boudoukh et al. 2007 & Accounting & 35.9 \\ 
  Asset Tangibility & Hahn and Lee 2009 & Accounting & 34.9 \\ 
  Advertising Expense & Chan et al. 2001 & Accounting & 31.3 \\ 
  IPO and Age & Ritter 1991 & Other & 14.2 \\ 
  Earnings Surprise Streak & Loh and Warachka 2012 & Accounting & 14.1 \\ 
  Conglomerate Return & Cohen and Lou 2012 & Price & 8.4 \\ 
  Customer Momentum & Cohen and Frazzini 2008 & Other & 4.9 \\ 
  Inst Own for High Short Int & Asquith Pathak and Ritter 2005 & Other & 0.2 \\ 
   \bottomrule
\end{tabular}

\end{center} 
\end{table}
\clearpage\pagebreak{}
\begin{table}
\caption{\textbf{The Missing Value Problem for Large-Scale Predictability Studies}}
\label{tab:miss-sum} \pdfbookmark{Table 2}{sum stat}

\begin{singlespace}
\noindent Data consists of 159 continuous predictors from Chen and
Zimmermann (2022) with at least 2 observations for common stocks listed
on major exchanges in every month between 1985 and 2021. Panel (a)
computes the \% of stocks with observations for individual predictors
and then computes percentiles across predictors. Panel (b) selects
the $J$ most-observed predictors and then computes the \% observed
among these using two methods: `Predictor-Stocks' sums predictor-stock
observations and then divides by $J\times$ the number of stocks.
`Stocks w/ all $J$' counts the number of stocks with all $J$ predictors
and divides by the number of stocks. \textbf{Interpretation:} Missing
values are a much larger problem when combining many predictors. When
combining 75 or more predictors, one must impute missing values or
drop most stocks.
\end{singlespace}
\begin{centering}
\vspace{0ex}
\par\end{centering}
\centering{}\setlength{\tabcolsep}{2ex} \small
\begin{center}
\begin{tabular}{crrrrrrrr} \toprule 
\multicolumn{9}{c}{Panel (a): Individual Predictors} \\ 
Predictor &   & \multicolumn{6}{c}{\% of Stocks With Predictor Data in June} &  \\ 
Percentile &   & 1970 & 1975 & 1980 & 1985 & 1990 & 1995 & 2000 \\ \cmidrule{1-1}\cmidrule{3-9}
\csname @@input\endcsname exhibits/pct_obs_one_signal_ptile.tex
\vspace{-3ex} \\
\bottomrule
\end{tabular}%
\end{center} 

\setlength{\tabcolsep}{1ex} 
\begin{center}
\begin{tabular}{rrrrrrrrrr} \toprule \multicolumn{10}{c}{Panel (b): Combining $J$ Predictors} \\
\midrule   &   & \multicolumn{2}{c}{June 1975} &   & \multicolumn{2}{c}{June 1985} &   & \multicolumn{2}{c}{June 1995} \\ 
\cmidrule{1-1}\cmidrule{3-4}\cmidrule{6-7}\cmidrule{9-10}  &   & \multicolumn{1}{l}{Predictor-} & \multicolumn{1}{l}{Stocks w/ $J$} &   & \multicolumn{1}{l}{Predictor-} & \multicolumn{1}{l}{Stocks w/ $J$} &   & \multicolumn{1}{l}{Predictor-} & \multicolumn{1}{l}{Stocks w/ $J$} \\ \multicolumn{1}{c}{$J$} &   & \multicolumn{1}{l}{Stocks w/} & \multicolumn{1}{l}{Predictors} &   & \multicolumn{1}{l}{Stocks w/} & \multicolumn{1}{l}{Predictors} &   & \multicolumn{1}{l}{Stocks w/} & \multicolumn{1}{l}{Predictors} \\   &   & \multicolumn{1}{l}{Data (\%)} & \multicolumn{1}{l}{(\%)} &   & \multicolumn{1}{l}{Data (\%)} & \multicolumn{1}{l}{(\%)} &   & \multicolumn{1}{l}{Data (\%)} & \multicolumn{1}{l}{(\%)} \\ 
\cmidrule{1-1}\cmidrule{3-4}\cmidrule{6-7}\cmidrule{9-10}
\csname @@input\endcsname exhibits/pct_obs_by_nsignal.tex
\bottomrule \end{tabular}%
\end{center} 
\end{table}
\pagebreak{}

\begin{table}
\caption{\textbf{Missing Data Effects on Single Predictor Strategies}}
\label{tab:one-signal} \pdfbookmark{Table 3}{single strats}

\begin{singlespace}
\noindent Strategies long the 500 stocks with the most positive predictors
and short the 500 stocks with the most negative predictors. `EM Imputed'
uses cross-sectional EM imputation (Section \ref{sec:method-EM})
and `Missing Dropped' drops stock-months with missing predictors.
For both methods, we drop predictor-months with less than 1000 stock-predictor
observations. \emph{`EM Improvement' }is `EM Imputed' minus `Missing
Dropped,' computed at the predictor level.\textbf{ Interpretation:}
Though EM generally improves performance, the typical effect is small,
and it comes at the cost of worse performance for many predictors.
\end{singlespace}
\begin{centering}
\vspace{0ex}
\par\end{centering}
\centering{}\setlength{\tabcolsep}{2ex} 
\begin{center}
\begin{tabular}{rlrr rrr} \toprule   &   & \multicolumn{5}{c}{Percentile Across Predictors} \\   &   & 10 & 25 & 50 & 75 & 90 \\ \midrule \multicolumn{7}{c}{Panel (a): Equal-Weighted} \\ \midrule 
\csname @@input\endcsname exhibits/one_signal_ew.tex
\midrule \\
\multicolumn{7}{c}{Panel (b): Value-Weighted} \\ \midrule
\csname @@input\endcsname exhibits/one_signal_vw.tex
\bottomrule 
\end{tabular}%
\end{center} 
\end{table}

\clearpage\pagebreak{}
\begin{table}
\caption{\textbf{Performance Across Several Imputation and Return Forecast
Methods}}
\label{tab:imp-fore-grid} \pdfbookmark{Table 4}{imp-fore-grid}

\begin{singlespace}
\noindent We impute missing values, forecast returns, and then form
long-short decile portfolios recursively each June between 1995-2021.
`Mean' imputes with cross-sectional means and `EM' is cross-sectional
EM (Section \ref{sec:method-EM}). Other imputations are defined in
Section \ref{sec:broad-impmethod}. `PCR' and `sPCR' run OLS on PCs
or scaled PCs (Section \ref{sec:result-pcr}). Other forecasts are
defined in Section \ref{sec:broad-foremethod}.\textbf{ }Tuning is
described in\textbf{ }Section \ref{sec:broad-oostune}. Forecasts
are either estimated within market equity groups (\emph{`}By Size\emph{')
}or using all stocks (`\textit{Pool}')\emph{.}\textbf{ Interpretation:}
Across many forecasting tests, mean imputation performs well compared
to other imputations, including imputations that incorporate time-series
information and ones that regularize using a factor structure (see
also Figure \ref{fig:bllp}). Mean imputation can even outperform
EM imputation if the forecasts ignore the size dependence of predictability.
\end{singlespace}
\begin{centering}
\vspace{-1ex}
\par\end{centering}
\begin{centering}
\setlength{\tabcolsep}{0.3ex} 
\begin{center} \footnotesize
\begin{tabular}{lrrrr rrrrr rrrrr rrr} \toprule \multicolumn{18}{c}{Panel (a): Equal-Weighted Mean Return (\% Annualized)} \\ \midrule   & \multicolumn{17}{c}{Forecasting Methods} \\ \cmidrule{2-18}  & \multicolumn{8}{c}{\textit{Linear}} &   & \multicolumn{8}{c}{\textit{Non-Linear}} \\ \cmidrule{2-9}\cmidrule{11-18}  & \multicolumn{2}{c}{OLS} &   & \multicolumn{2}{c}{PCR} &   & \multicolumn{2}{c}{sPCR} &   & \multicolumn{2}{c}{GBRT} &   & \multicolumn{2}{c}{NN1} &   & \multicolumn{2}{c}{NN3} \\   & \multicolumn{1}{l}{By Size} & \multicolumn{1}{l}{\textit{Pool}} &   & \multicolumn{1}{l}{By Size} & \multicolumn{1}{l}{\textit{Pool}} &   & \multicolumn{1}{l}{By Size} & \multicolumn{1}{l}{\textit{Pool}} &   & \multicolumn{1}{l}{By Size} & \multicolumn{1}{l}{\textit{Pool}} &   & \multicolumn{1}{l}{By Size} & \multicolumn{1}{l}{\textit{Pool}} &   & \multicolumn{1}{l}{By Size} & \multicolumn{1}{l}{\textit{Pool}} \\ \cmidrule{2-3}\cmidrule{5-6}\cmidrule{8-9}\cmidrule{11-12}\cmidrule{14-15}\cmidrule{17-18}
\csname @@input\endcsname exhibits/fcast-ew.tex
\vspace{-3ex} \\
\bottomrule
\end{tabular}%
\end{center} 
\par\end{centering}
\centering{}\setlength{\tabcolsep}{0.3ex} 
\begin{center} \footnotesize
\begin{tabular}{lrrrr rrrrr rrrrr rrr} \toprule \multicolumn{18}{c}{Panel (b): Value-Weighted Mean Return (\% Annualized)} \\ \midrule   & \multicolumn{17}{c}{Forecasting Methods} \\ \cmidrule{2-18}  & \multicolumn{8}{c}{\textit{Linear}} &   & \multicolumn{8}{c}{\textit{Non-Linear}} \\ \cmidrule{2-9}\cmidrule{11-18}  & \multicolumn{2}{c}{OLS} &   & \multicolumn{2}{c}{PCR} &   & \multicolumn{2}{c}{sPCR} &   & \multicolumn{2}{c}{GBRT} &   & \multicolumn{2}{c}{NN1} &   & \multicolumn{2}{c}{NN3} \\   & \multicolumn{1}{l}{By Size} & \multicolumn{1}{l}{\textit{Pool}} &   & \multicolumn{1}{l}{By Size} & \multicolumn{1}{l}{\textit{Pool}} &   & \multicolumn{1}{l}{By Size} & \multicolumn{1}{l}{\textit{Pool}} &   & \multicolumn{1}{l}{By Size} & \multicolumn{1}{l}{\textit{Pool}} &   & \multicolumn{1}{l}{By Size} & \multicolumn{1}{l}{\textit{Pool}} &   & \multicolumn{1}{l}{By Size} & \multicolumn{1}{l}{\textit{Pool}} \\ \cmidrule{2-3}\cmidrule{5-6}\cmidrule{8-9}\cmidrule{11-12}\cmidrule{14-15}\cmidrule{17-18}
\csname @@input\endcsname exhibits/fcast-vw.tex
\vspace{-3ex} \\
\bottomrule
\end{tabular}%
\end{center} 
\end{table}
\clearpage\pagebreak{}
\begin{figure}[h!]
\caption{\textbf{Missingness Map: June 1990.}}
\label{fig:missmap} \pdfbookmark{Figure 1}{missmap} Vertical axis
represents 159 predictors and horizontal axis represents 6,000 common
stocks traded on major exchanges. Shading indicates months since the
last observation. Lightest shade is currently observed and the darkest
indicates the predictor-stock has never been observed. Predictors
are sorted by the type of data they focus on (according to the CZ
documentation). \textbf{Interpretation: }When combining 159 predictors,
a stock can only be used without imputation if it has values for all
159 predictors. No stocks satisfy this requirement. The complexity
of missingness suggests assuming \citet{rubin1976inference} ignorability
is required. Missingness occurs in blocks organized around data type
and time. There is very little time series information about missing
values.

\vspace{0.15in}

\centering
\includegraphics[width=0.9\textwidth]{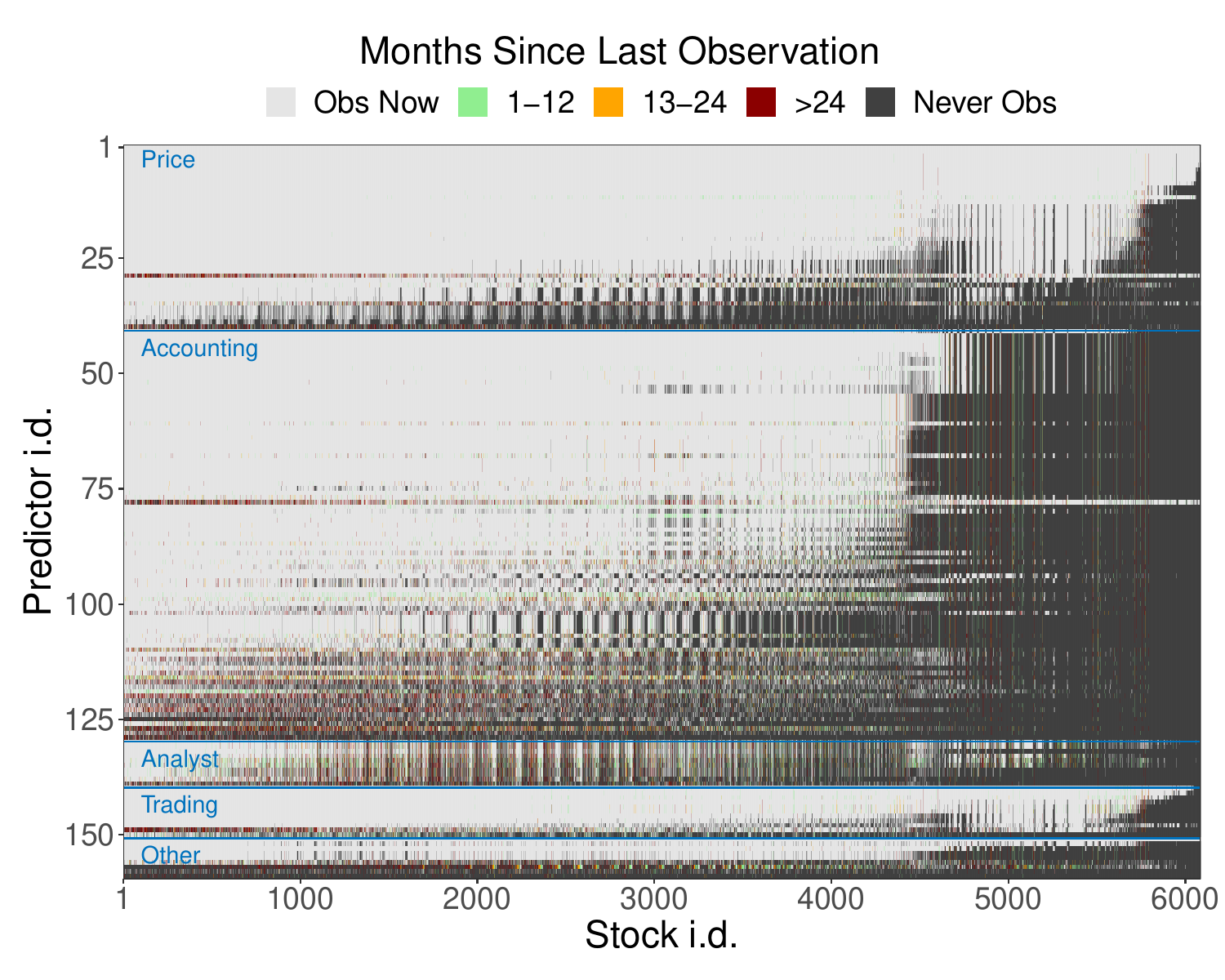}      
\end{figure}
\clearpage\pagebreak{}
\begin{figure}[h!]
\caption{\textbf{Missing Data Effects in 159-Predictor Strategies using PCR}}
\label{fig:pca-bysize} \pdfbookmark{Figure 2}{pca-by-size} Each
month, we forecast next month's returns by running PCR on the past
10 years of data by size group (micro, small, big). Strategies go
long-short stocks in the extreme deciles of predicted returns. `EM
Algo' imputes missing values with cross-sectional EM (Section \ref{sec:method-EM}).
`Simple Mean' imputes with cross-sectional means (Section \ref{sec:method-simplemean}).\textbf{
Interpretation: }EM and simple mean imputation lead to similar inferences
about expected returns and Sharpe ratios.

\vspace{0.15in}

\centering 
\subfloat[Mean Returns]{\includegraphics[width=0.9\textwidth]{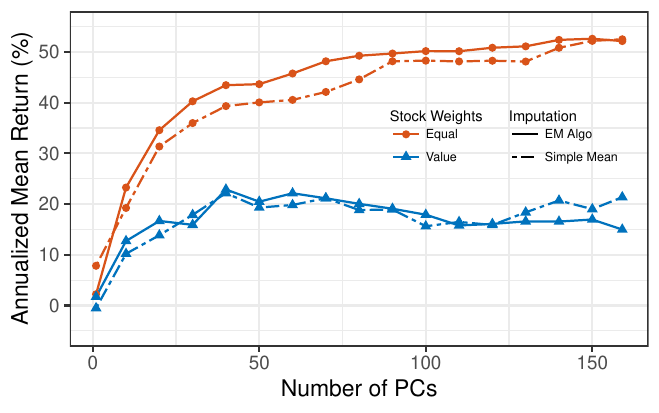}}\\
\subfloat[Sharpe Ratios]{\includegraphics[width=0.9\textwidth]{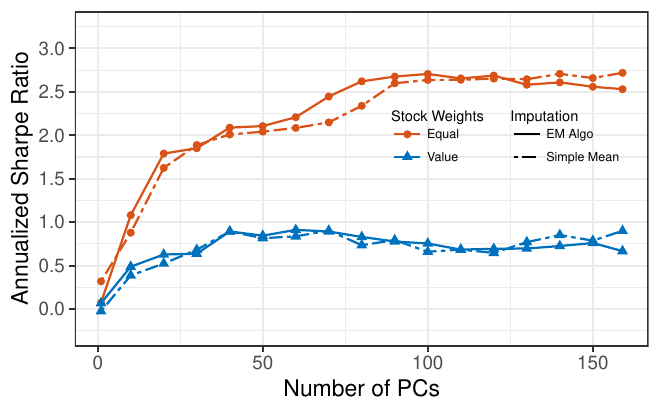}}   
\end{figure}
\clearpage\pagebreak{}
\begin{figure}[h!]
\caption{\textbf{Empirical Distribution of Correlations}}
\label{fig:cor} \pdfbookmark{Figure 3}{corr-dist} All predictors
are transformed using an extended Box-Cox algorithm to have approximately
normal marginal distributions at the predictor level (Section \ref{sec:method-boxcox}).
`Observed' uses all available cases at the predictor-pair level (a.k.a.
pairwise-complete). `EM Algo' uses the cross-sectional EM-estimated
covariance matrix\textbf{. Interpretation:} The correlations cluster
near zero, implying that imputation slopes are close to zero (Equation
(\ref{eq:em-slope})), and thus simple mean imputation should be similar
to EM imputation. \vspace{0.15in}

\centering 
\subfloat[June 1990]{\includegraphics[width=0.45\textwidth]{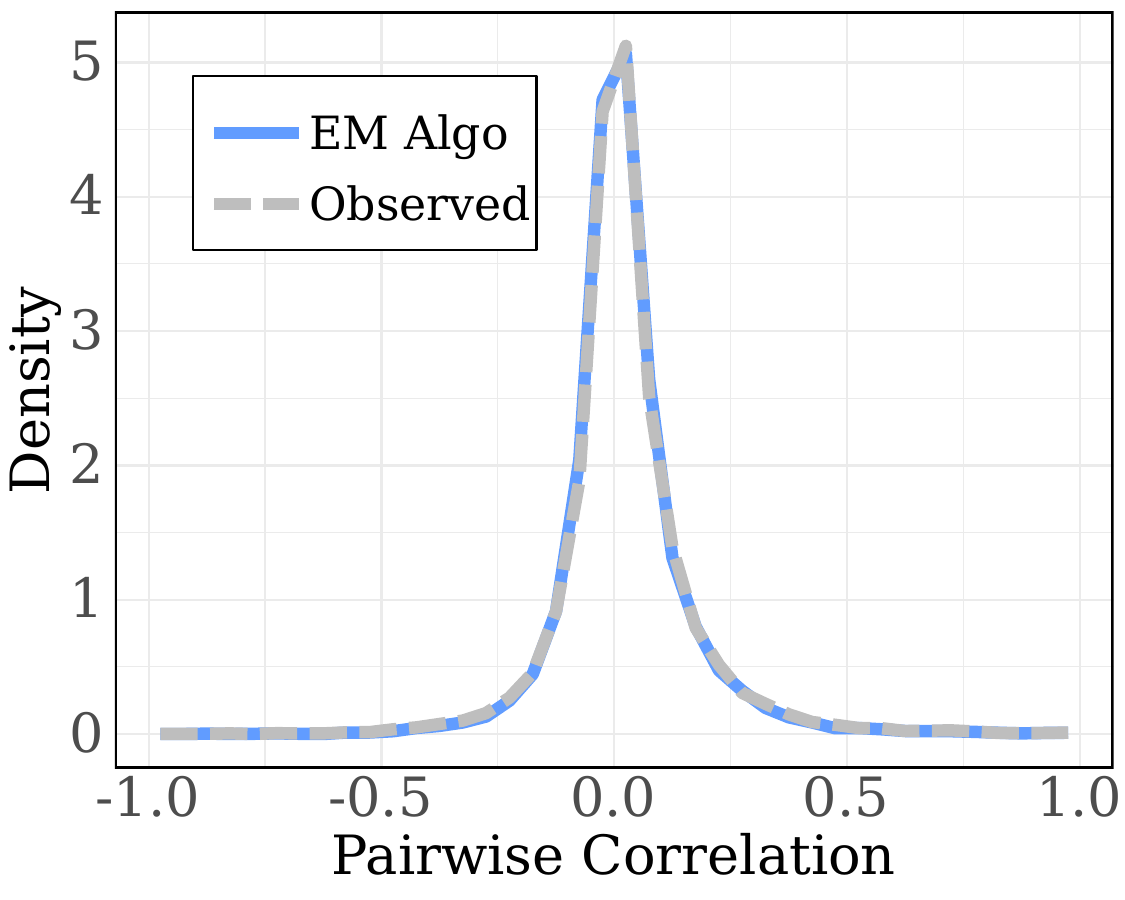}
\includegraphics[width=0.45\textwidth]{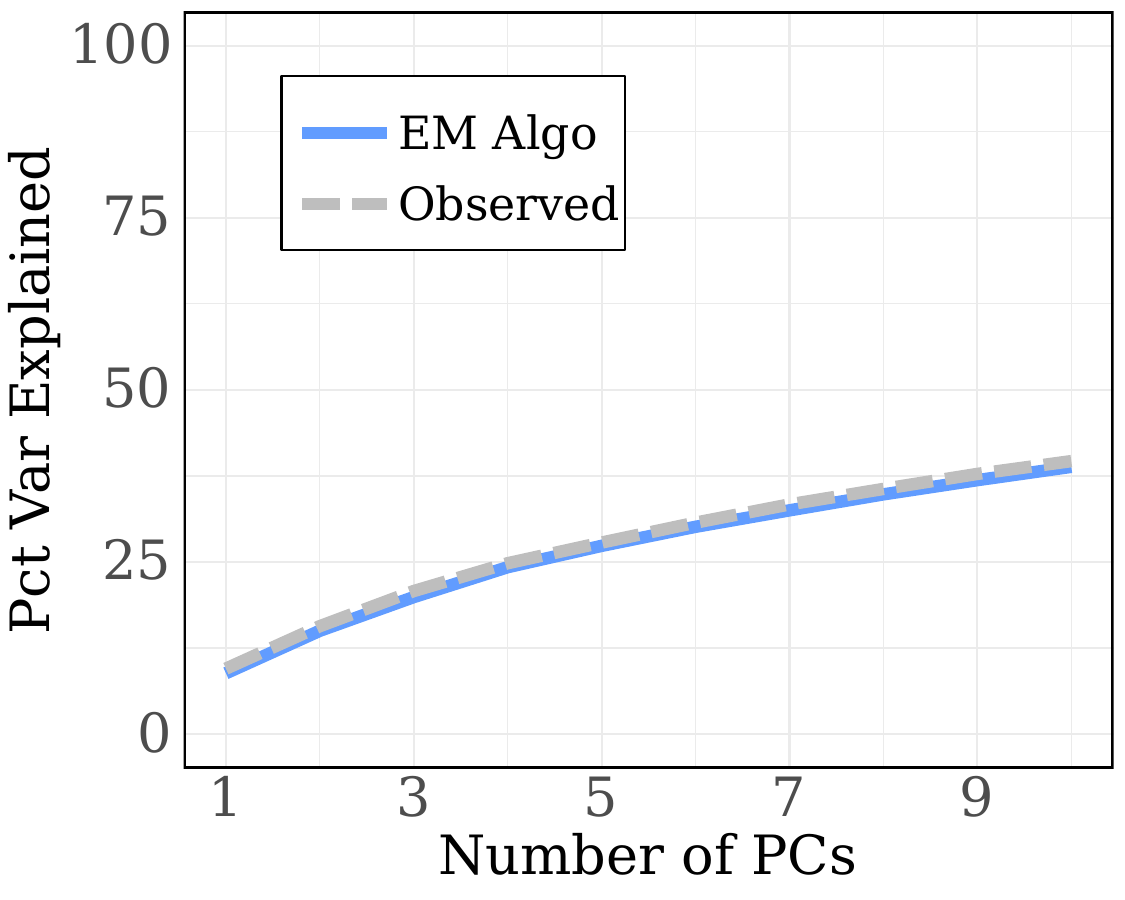}}\\
\subfloat[June 2000]{\includegraphics[width=0.45\textwidth]{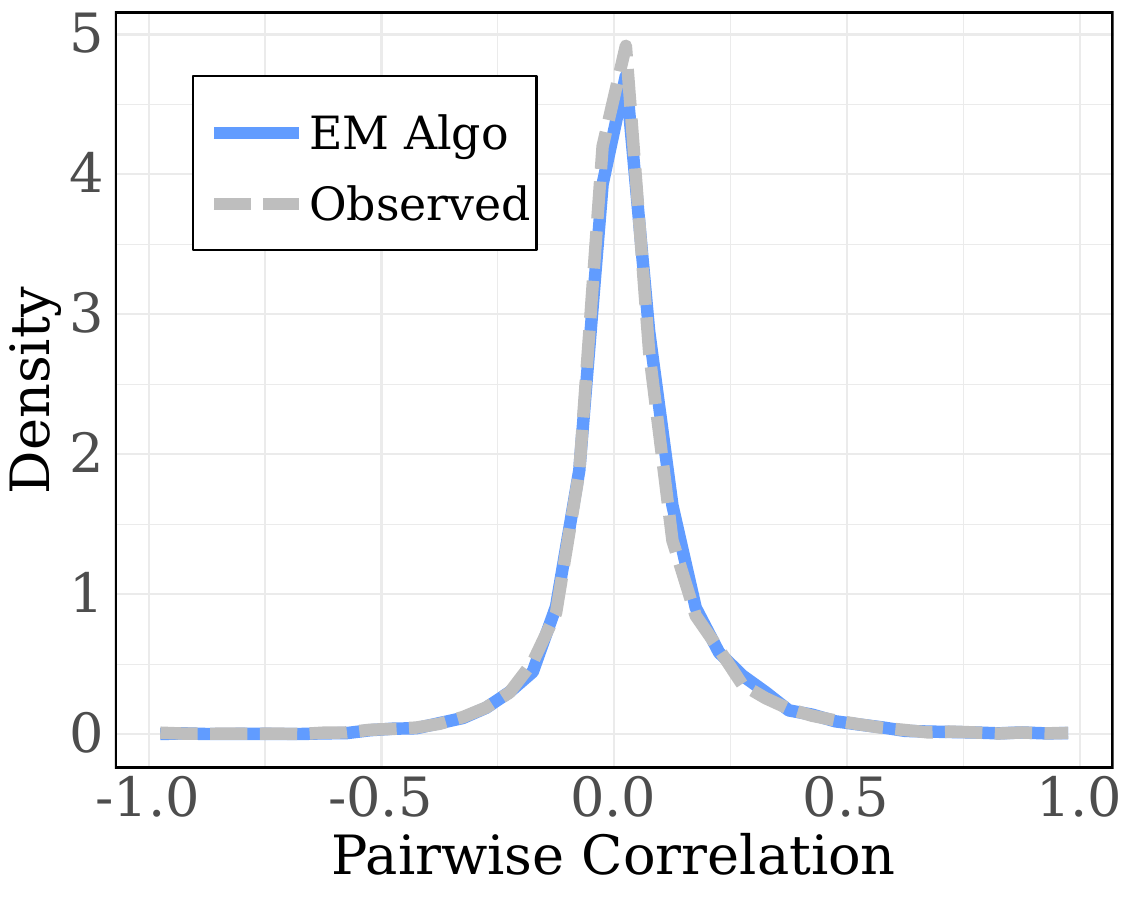}
\includegraphics[width=0.45\textwidth]{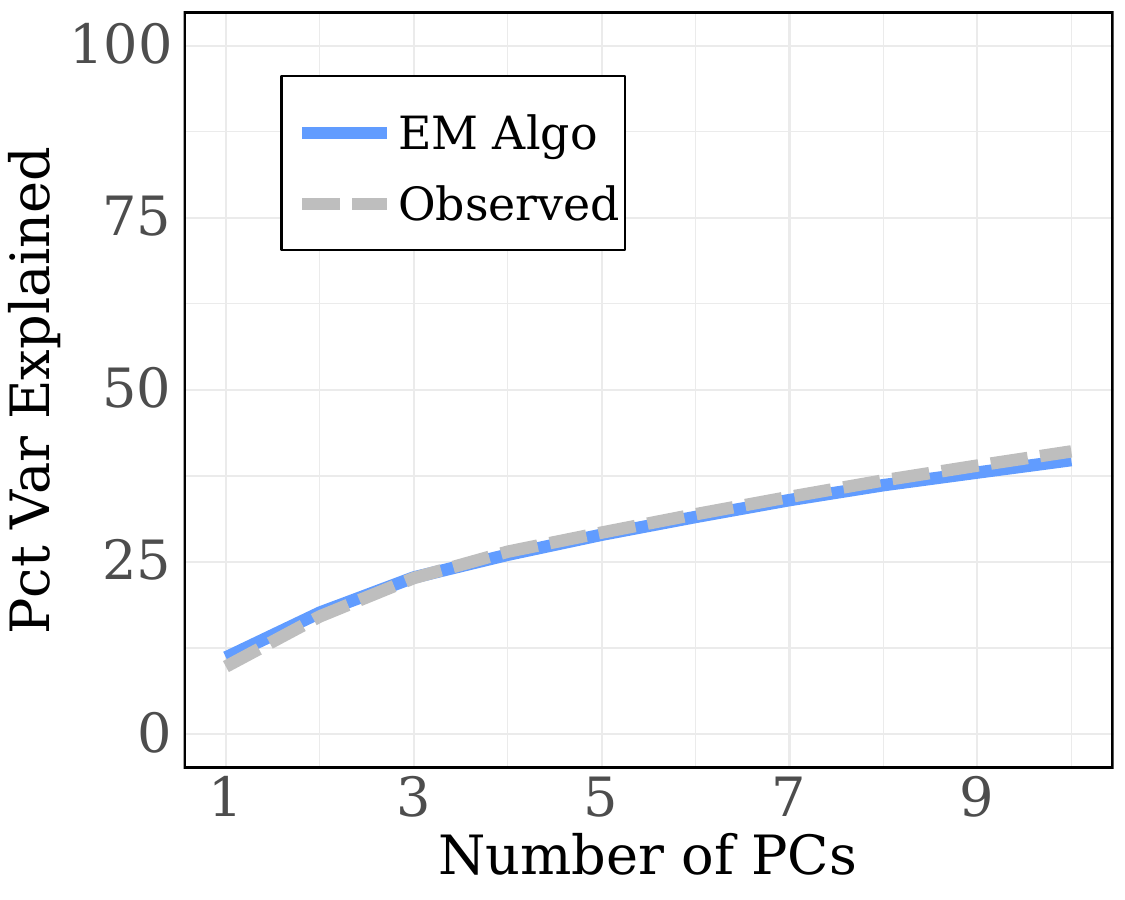}
}   
\\ 
\subfloat[June 2010]{
\includegraphics[width=0.45\textwidth]{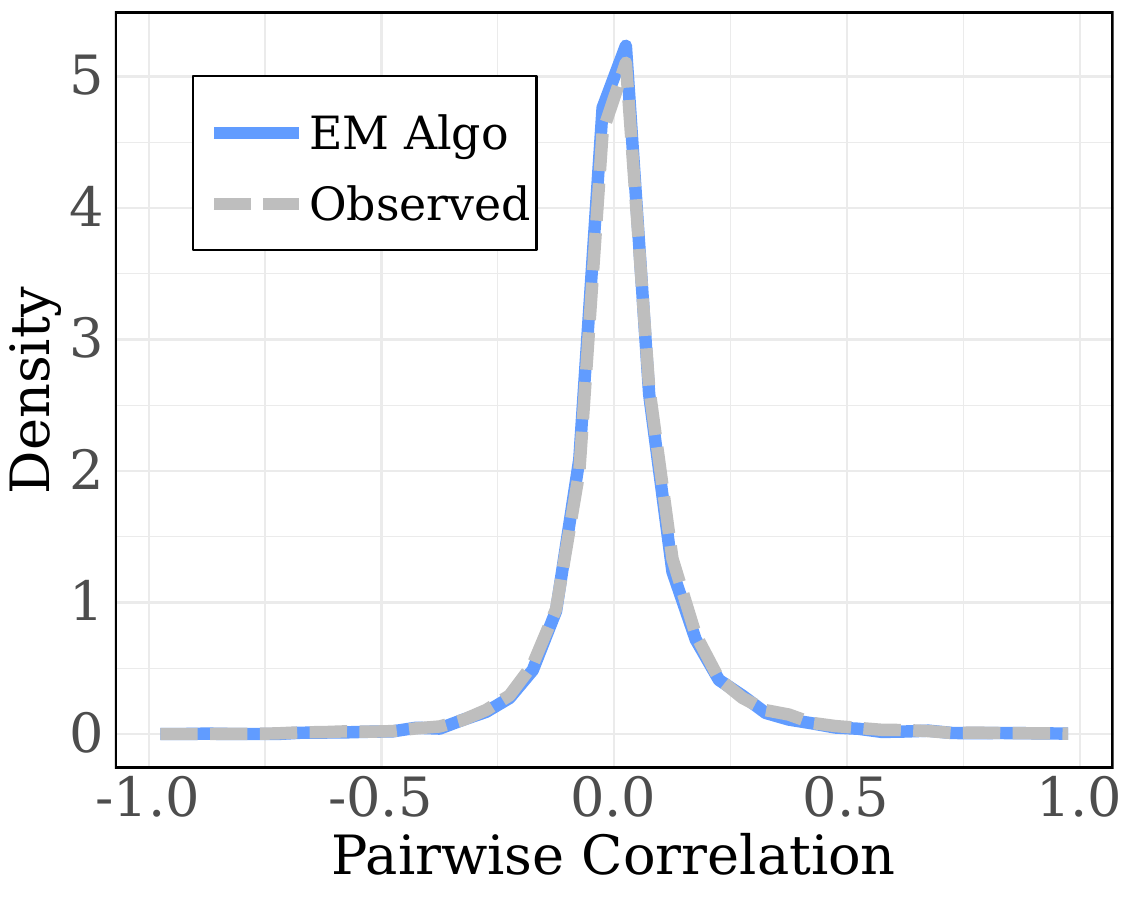}
\includegraphics[width=0.45\textwidth]{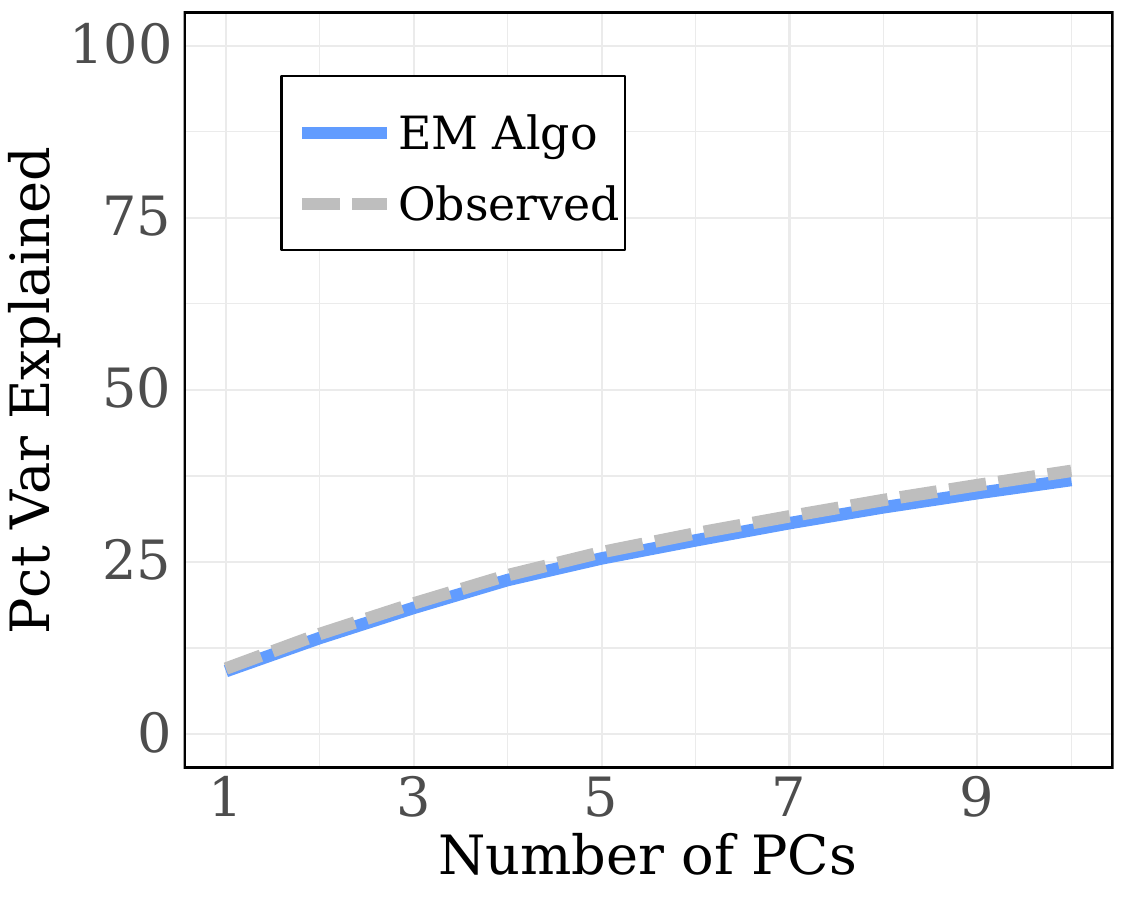}
}   
\end{figure}

\clearpage\pagebreak{}
\begin{figure}[h!]
\caption{\textbf{Imputation Errors in a Random Masking Exercise}}
\label{fig:impute-err} \pdfbookmark{Figure 4}{impute-err} We measure
out-of-sample imputation errors a la cross-validation as follows:
For each cross-section, we randomly divide the stock-predictor observations
into 10 groups. We mask one of the groups, impute with cross-sectional
EM, and find the imputation errors for the masked stock-predictors.
Masking and imputation is repeated for each of the 10 groups. We then
calculate RMSE by market equity decile.\textbf{ Interpretation: }Imputations
errors are very large among small stocks, nearly as large as the errors
from mean imputation, which is about 1.0 by construction.

\vspace{0.15in}

\centering 
\includegraphics[width=0.9\textwidth]{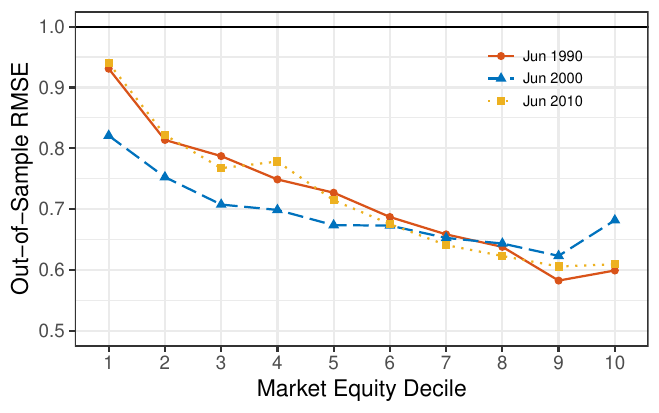}
\end{figure}
\clearpage\pagebreak{}
\begin{figure}[h!]
\caption{\textbf{Missing Data Effects in PCR Strategies Using Pooled Forecasts}}
\label{fig:pca-pooled} \pdfbookmark{Figure 5}{pca-pooled} Like Figure
\ref{fig:pca-bysize}, we form portfolios on PCR return forecasts,
but now we forecast all stocks together instead of forecasting micro,
small, and big stocks separately. `EM Algo' imputes missing values
with the cross-sectional EM algorithm (Section \ref{sec:method-EM}).
`Simple Mean' imputes with cross-sectional means (Section \ref{sec:method-simplemean}).\textbf{
Interpretation: }Large imputation errors in small stocks (Figure \ref{fig:impute-err})
lead to poor performance in value-weighted strategies, if micro and
big stock returns are assumed to have the same structure. EM introduces
estimation noise, which can lead to underperformance relative to mean
imputation, particularly if forecasts are not carefully designed.

\vspace{0.15in}

\centering 
\subfloat[Mean Returns]{\includegraphics[width=0.9\textwidth]{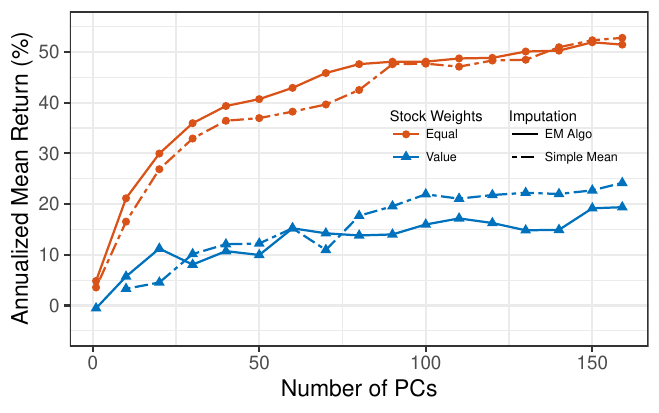}}\\
\subfloat[Sharpe Ratios]{\includegraphics[width=0.9\textwidth]{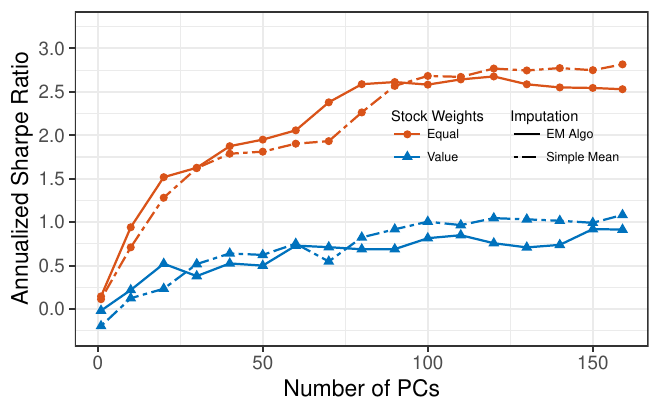}}   
\end{figure}

\clearpage\pagebreak{}
\begin{figure}[h!]
\caption{\textbf{Missing Data Effects on 159-Predictor Strategies using Scaled-PCA}}
\label{fig:spca-bysize} \pdfbookmark{Figure 6}{spca-bysize} Like
Figure \ref{fig:pca-bysize}, we form portfolios on PCR return forecasts,
but now we construct PCs using \citet{huang2022scaled}'s scaled-PCA
instead of standard PCA. Panel (b) zooms in on the first 25 PCs. \textbf{Interpretation:
}PCA significantly overstates the dimensionality of expected returns.
Scaled PCA requires far fewer PCs to capture the potential returns,
especially if EM is used. The dimensionality of expected returns is
subtle and depends on multiple measurement decisions. Mean imputation
still does a good job capturing the potential mean returns.

\vspace{0.15in}

\centering 
\subfloat[Mean Return: All PCs]{\includegraphics[width=0.9\textwidth]{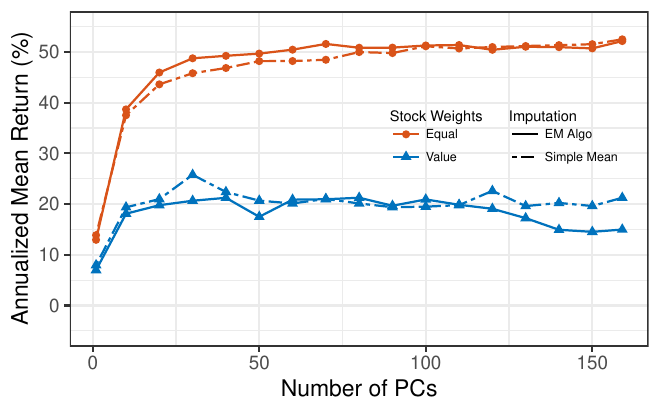}}\\
\subfloat[Mean Returns: PCs 1-25]{\includegraphics[width=0.9\textwidth]{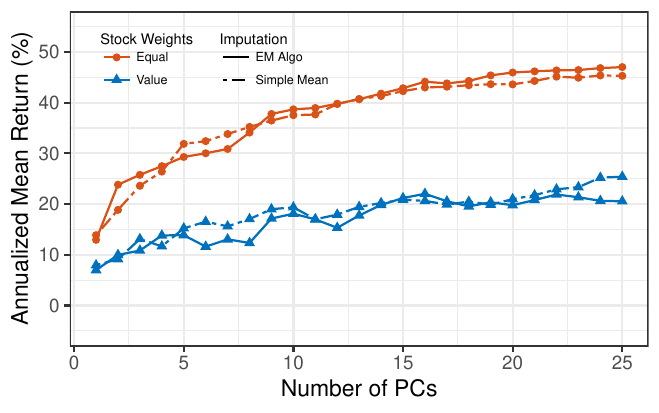}}   
\end{figure}

\end{document}